\documentclass[
 reprint,
 amsmath,amssymb,
 aps, prx,
floatfix,longbibliography,
]{revtex4-2}

\usepackage{graphicx} 
\usepackage{dcolumn} 
\usepackage{sidecap}
\usepackage{float}
\usepackage{physics}
\usepackage{bm}  

\usepackage[svgnames]{xcolor}
\usepackage[bookmarks=false,linkcolor=blue,urlcolor=blue,colorlinks,citecolor=blue]{hyperref}

\usepackage{lipsum}
\usepackage{booktabs}

\newcommand{\FNN}{self-attention NN }

\begin{document}

\title{Is attention all you need to solve the correlated electron problem?}

\author{Max Geier}
\thanks{These two authors contributed equally.}
\affiliation{Department of Physics, Massachusetts Institute of Technology, Cambridge, MA 02139, USA}

\author{Khachatur Nazaryan}
\thanks{These two authors contributed equally.}
\affiliation{Department of Physics, Massachusetts Institute of Technology, Cambridge, MA 02139, USA}

\author{Timothy Zaklama}
\affiliation{Department of Physics, Massachusetts Institute of Technology, Cambridge, MA 02139, USA}

\author{Liang Fu}
\affiliation{Department of Physics, Massachusetts Institute of Technology, Cambridge, MA 02139, USA}

\date{\today} 

\begin{abstract}
The attention mechanism has transformed artificial intelligence research by its ability to learn relations between objects. 
In this work, we explore how a many-body wavefunction ansatz constructed from a large-parameter self-attention neural network can be used to solve the interacting electron problem in solids. 
By a systematic neural-network variational Monte Carlo study on a moir\'e quantum material, we demonstrate that the self-attention ansatz provides an accurate and efficient solution without human bias.
Moreover, our numerical study finds that the required number of variational parameters scales roughly as $N^2$ with the number of electrons, which opens a path towards efficient large-scale simulations.
\end{abstract}

\maketitle

\section{Introduction}

Solving the many-electron Schr\"odinger equation for solids is an exceedingly difficult problem due to the exponential growth of the Hilbert space dimension. 
Various techniques based on the variational principle have long been developed to approximate the ground state of interacting electron systems using trial wavefunctions.  
The Hartree-Fock method based on independent electron approximation \cite{Hartree1928Jan,Fock1930Jan} usually captures 99\% of the total energy \cite{Foulkes2001Jan}, but misses electron correlation effects, the driving force behind fascinating quantum phenomena such as high-temperature superconductivity and the fractional quantum Hall effect.

Recently, neural network (NN) based variational Monte Carlo (VMC) has been developed for solving the many-electron problem with high accuracy   \cite{Carleo2017Feb,Schutt2017Dec,Carleo2019Jul,Luo_backflow,Choo2020May,PauliNet2020, Pfau2020Sep,vonGlehn2022Nov, viteritti2023transformer}. 
Compared to human-designed trial wavefunctions, neural network wavefunctions contain a large number of parameters, has enormous representation power, and can be optimized efficiently. NN-VMC has been shown to be highly accurate in calculating the ground state energy of interacting electrons in atoms and molecules \cite{PauliNet2020, Pfau2020Sep, vonGlehn2022Nov,Li2023JulForwardLaplacian,Gao2023Jul}, lattice models \cite{Carleo2019Jul,Moreno2022,viteritti2023transformer,Luo2023Mar,ChenJ1J22024},
uniform electron gas \cite{FerminetSolid2023, Wilson2023Jun, Luo2023,Kim2024May,Smith2024,pescia2023message,sobral2024physics},
moire semiconductors \cite{Li2024,Luo2024},
and fractional quantum Hall liquids \cite{Teng2024Nov,Qian2024}.

Despite the rapid progress, two important questions remain open.  
First, a number of NN architectures have so far been introduced and used to study different many-electron problems.  
Is there any hope of finding a unifying architecture that applies to a wide range of interacting electron systems?
Second, it is essential to assess the finite size effect in numerical simulations of solid state systems.  
How does the performance of the neural ansatz change as the system size increases?

In this work, we present a general NN-VMC method to solve many-electron problems in solids, where electron correlations in the NN ansatz are {\it entirely} produced from the self-attention mechanism.
The attention mechanism was originally introduced in the context of large language models to learn relations between words \cite{transformer}. 
Here, the attention mechanism is employed to identify and quantify 
how electrons influence each other and how such influence affects their individual orbitals. 
This enable the construction of NN wavefunctions from Slater determinants of generalized orbitals that depend on the configuration of all electrons \cite{vonGlehn2022Nov}.  
By minimizing the energy using Monte Carlo and neural network techniques \cite{Foulkes2001Jan}, variational energy and wavefunction for the many-electron ground state are obtained.    

The performance of our attention based NN ansatz is evaluated for interacting electrons in semiconductor moir\'e heterobilayers, such as WSe$_2$/WS$_2$. 
This moir\'e platform hosts a fascinating variety of correlated electronic states, including Mott insulators \cite{regan2020mott, tangSimulationHubbardModel2020a,Arsenault2024Mar}, generalized Wigner crystals \cite{li2021imaging}, and strongly correlated Fermi liquids \cite{Zhao2023}. 
Here, doped electrons reside on one semiconductor layer and experience a moir\'e potential. The Hamiltonian thus takes the form of a two-dimensional Coulomb electron gas in a periodic potential \cite{Wu2018}: 
\begin{align}
    H & = H_0 + H_{ee} \nonumber\\
    & = \sum_i \left( 
-\frac{1}{2}\nabla_i^2 + V(\mathbf{r}_i) 
\right) 
+ \frac{1}{2} \sum_i \sum_{i \neq j} \frac{1}{|\mathbf{r}_i - \mathbf{r}_j|}, 
\label{eq:system-hamiltonian}
\end{align}
where $V(\mathbf{r}) = -2V_0 \sum_{j=1}^3 \cos(\mathbf{g}_j \cdot \mathbf{r} + \varphi)$ is the moir\'e potential with reciprocal lattice vectors $\mathbf{g}_j = \frac{4\pi}{\sqrt{3}a_M} (\cos \frac{2\pi j}{3}, \sin \frac{2\pi j}{3})$, moir\'e lattice constant $a_M$, and $\varphi$ controls the shape of the moir\'e potential. Despite its conceptual simplicity, this Hamiltonian exhibits a variety of electron phases that emerge from the interplay between kinetic energy, moir\'e potential and Coulomb interaction \cite{zhang2020density, reddy2023artificial}.

First, for small system size, we benchmark our NN results with band-projected exact diagonalization. Remarkably, the NN energies are found to be lower even when five bands are included in the exact diagonalization. 
Next, we assess the performance of the NN wavefunction as the system size increases. 
Specifically, we study how the required number of parameters $N_{\rm par}$ scales with the number of electrons $N$,
and numerically discover a scaling law $N_{\rm par} \propto N^\alpha$ with a small exponent $\alpha \approx 2$.
These findings suggests that self-attention NN ansatz is an accurate and efficient method in solving large-scale interacting electron problems. 

We note that self-attention NN wavefunctions were first introduced in Ref.~\cite{vonGlehn2022Nov} to solve quantum chemistry problems, reaching state-of-the-art accuracy.
The recent work Ref.~\cite{Teng2024Nov} demonstrated that the self-attention ansatz accurately describes the fractional quantum Hall ground state, achieving a lower ground state energy than Landau-level projected exact diagonalization. 
The remarkable success of the self-attention NN wavefunction in atoms, molecules, electron gas, and now moir\'e materials---taken as a whole---suggests that self-attention is a key ingredient for a unifying solution to the correlated electron problem. 

In the remainder of the article, we detail the self-attention neural network wavefunction ansatz and present its benchmark with self-consistent Hartree-Fock and band-projected exact diagonalization. 
First, in Sec.~\ref{sec:traditional_methods} we summarize the traditional methods of Hartree-Fock and band-projected exact diagonalization for comparison. Then, 
Sec.~\ref{sec:NN} describes the construction of the self-attention neural network ansatz. 
Sec.~\ref{sec:VMC} summarizes principal ideas of the variational Monte Carlo method applied to optimize the wavefunction ansatz. Finally, our numerical results and benchmarks are presented in Sec.~\ref{sec:results}.

\section{Traditional Numerical methods}
\label{sec:traditional_methods}
In this section, we present Hartree-Fock and band-projected exact diagonalization solutions of the interacting Hamiltonian for TMD heterobilayers. Results from these standard numerical methods will provide the benchmark for NN-VMC in the following sections.       

Throughout this work we study periodic solids. Numerical simulation of a periodic solid requires to divide the space into periodic supercells of finite size and require periodic boundary conditions on the wavefunction.
With periodic boundary conditions, the Coulomb interaction needs to be carefully calculated to account for interactions of particles within the supercell with all their images in the other supercells, see App.~\ref{app:ewald} for details. 

\subsection{Hartree-Fock}

The Hartree-Fock approximation to the ground state of an interacting fermionic system is obtained by minimizing the energy over the space of single Slater determinant states,
\begin{equation}
    \Psi^{\rm HF}({\bm r}_1, ..., {\bm r}_N) = \frac{1}{\sqrt{N !}}\det \begin{pmatrix} \phi_1(\bm r_1) & ... & \phi_N(\bm r_1) \\ 
    \vdots & \ddots & \vdots \\
    \phi_1(\bm r_N) & ... & \phi_N(\bm r_N)\end{pmatrix}
    \label{eq:HF-SlaterDet}
\end{equation}
This wavefunction ansatz captures the quantum-mechanical "exchange" effect by incorporating the Pauli exclusion principle, but neglects correlations arising from electron-electron interactions. Single Slater determinant states have a simple structure that enables efficient computation of observables using Wick's theorem. At the same time, this structure enables a self-consistent determination of the optimal one-body orbitals $\phi_j(\bm r)$ underlying the Slater determinant wavefunction.

By construction, the energy of the Hartree-Fock ground state is the energy of the best wavefunction constructed from independent orbitals. This motivates the definition of \emph{correlation energy} as the energy difference between the true ground state and the Hartree-Fock ground state
\begin{equation}
    E_{\rm corr} = E_{\rm GS} - \langle \Psi^{\rm HF} | \hat{H}| \Psi^{\rm HF} \rangle
    \label{eq:methods-correlation-energy}
\end{equation}
In molecular systems and most solid-state systems, the correlation energy typically is only about one percent of the ground state energy \cite{Foulkes2001Jan}. However, it plays an essential role in driving  various quantum phases of matter ranging from superconductivity to fractional quantum Hall states.

\subsection{Band-projected exact diagonalization}

One approach to capture electron-electron correlations is to construct a finite set of Slater determinant wavefunctions and exactly diagonalize the full Hamiltonian projected onto this finite subspace of the Hilbert space. This approach is variational because the restricted choice of Slater determinants forms a variational wavefunction ansatz. The obtained ground state energy is an upper bound on the true ground state energy, because inclusion of additional Slater determinants can further lower the energy. 

A good choice of Slater determinants spanning the variational subspace is obtained by diagonalizing the non-interacting part of the Hamiltonian $H_0$, i.e. the first term in Eq.~\eqref{eq:system-hamiltonian}, and keeping only the lowest energy eigenstates up to a cutoff. For a periodic system, the single-particle eigenstates are Bloch states labeled by the Bloch momentum $\bm k$ and band index $n$. Projecting onto the the subspace of lowest $N_{\rm bands}$ bands, the Hamiltonian matrix elements between Slater determinants is written in second quantization,
\begin{align}
    \mathbf{\tilde{H}}& = \sum_{n=1}^{N_{\rm bands}}\sum_{\mathbf{k},\sigma}\varepsilon_{\mathbf{k}\sigma}^nc_{\mathbf{k}\sigma n}^\dagger c_{\mathbf{k}\sigma n} \  + \frac{1}{2} \sum_i^N \xi_{\rm M}\nonumber \\
    & \ + \frac{1}{2} \sum_{\substack{n,m,\\n',m'=1}}^{N_{\rm bands}} \sum_{\substack{\mathbf{k'}\mathbf{p'}\mathbf{k}\mathbf{p}\\ \sigma\sigma'}} V^{nmn'm'}_{\mathbf{k'}\mathbf{p'}\mathbf{k},\mathbf{p};\sigma\sigma'} c_{\mathbf{k'}\sigma n}^\dagger c_{\mathbf{p'}\sigma' m}^\dagger c_{\mathbf{p}\sigma' m'} c_{\mathbf{k}\sigma n'}
    \label{eq:methods-ED-H}
\end{align}
where $c_{\mathbf{k},\sigma,n}^\dagger$ creates a Bloch state in the $n$'th band at crystal momentum $\mathbf{k}$ and spin $\sigma$ with corresponding single-particle energy $\varepsilon_{\mathbf{k}\sigma}^n$, and $V_{\mathbf{k'}\mathbf{p'}\mathbf{k},\mathbf{p};\sigma\sigma'}^{nmn'm'} \equiv \langle \mathbf{k'},\sigma, n;\mathbf{p'},\sigma',m|\hat{V}|\mathbf{k},\sigma,n';\mathbf{p},\sigma',m' \rangle$ are the corresponding matrix elements of the Coulomb interaction projected onto the band basis. Eq.~\eqref{eq:methods-ED-H} contains the Madelung energy $N \xi_{\rm M}/2$ that describes the interaction of charges in the periodic supercell with their own images in other supercells, see App.~\ref{app:ewald} for details.

For a small size system, the band-projected Hamiltonian $\mathbf{\tilde{H}}$ can be numerically diagonalized to yield the ground state energy and wavefunction. The band-projected exact digonalization (BP-ED) method is accurate only if the interaction strength is small compared to the gap to excitations involving high-energy bands. 
However, as we shall show below, interaction-induced band mixing is substantial for realistic material parameters, and therefore it is necessary to include {\it multiple} low-lying bands to obtain quantitatively accurate results, which limits the practical application of BP-ED to very small system size.

\section{Neural Network Wavefunction}
\label{sec:NN}

\begin{figure}
    \centering
    \includegraphics[width=\linewidth]{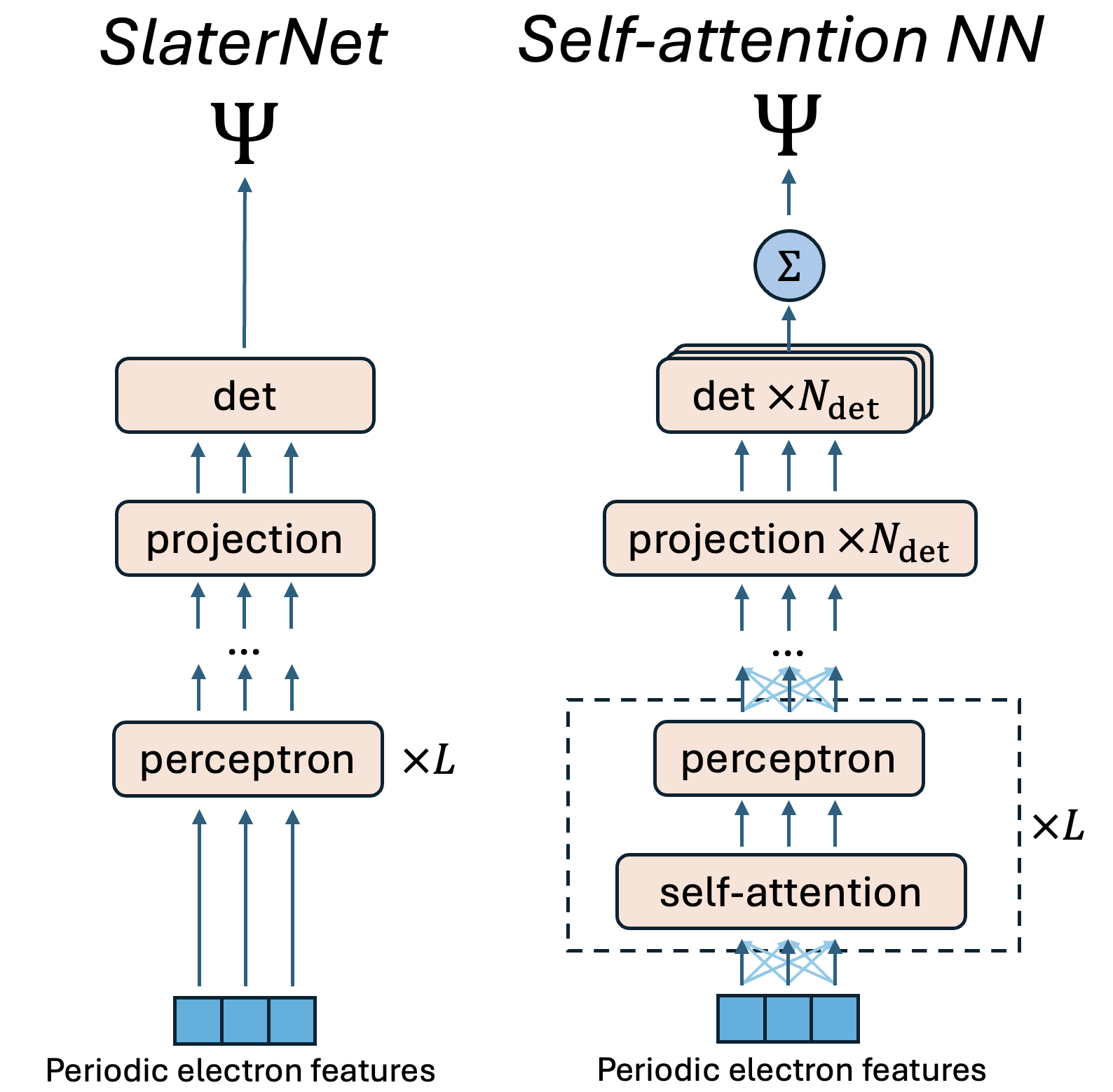}
    \caption{\textbf{Architecture of the neural network wavefunction ansatz.} {\em SlaterNet}: Multilayer perceptron neural network generates one-body orbitals to approximate general single Slater determinant wavefunctions. {\em Self-attention NN}: Self-attention neural network for solids based on Psiformer \cite{vonGlehn2022Nov} capturing correlations by 
    communicating mutual information between individual particle streams.}
    \label{fig:1}
\end{figure}

While Hartree-Fock and BP-ED studies have provided useful results on moir\'e semiconductors, they suffer from respective limitations. The Hartree-Fock method fails to capture electron correlation effect and cannot describe entangled states such as fractional Chern insulators, while BP-ED is severely limited by finite size effect and band truncation. 

Recently, a new type of variational method based on neural network wavefunctions has been developed to approximate the ground state of many-electron systems. Traditional variational wavefunctions for correlated systems are artfully designed by humans, tailor made for a given quantum phase, and usually contain only a few variational parameters. In contrast, thanks to their universal approximation capability, deep neural networks can be used to generate powerful wavefunction ansatz, which has a large number of variational parameters and can accurately represent distinct phases of matter in a unified way \cite{Hornik1989Jan,Pfau2020Sep,Kim2024MayAttentionToComplexity}.

In the following, we show how an expressive neural network wavefunction ansatz capturing correlations between particles can be constructed. 
First, in Sec.~\ref{sec:NN-onebody}, a universal approximation of single-particle orbitals is constructed by a deep feed-forward neural network. 
Second, Sec.~\ref{sec:NN-SlaterNet}, by generating a set of single-particle orbitals from the feed-forward neural network, a single Slater determinant wavefunction is obtained, which enables neural network implementation of unrestricted Hartree Fock calculations. 
Finally, Sec.~\ref{sec:NN-attn}, electron correlations are captured by mixing different electron streams in a permutation-equivariant way with self-attention mechanism. 
The resulting neural network wavefunction takes the form of generalized Slater determinants that are built from ``correlated orbitals'', and by construction, is antisymmetric in electron coordinates \cite{Pfau2020Sep}. 
This representation is in principle universal as all many-body fermionic wavefunctions can be written as a generalized Slater determinant when allowing for discontinuous functions \cite{Pfau2020Sep}.
Fig.~\ref{fig:1} shows schematically and compares the network architectures representing general single-Slater determinant wavefunctions, and its extension by self-attention capturing correlations.

\subsection{Deep neural network for one-body orbitals}
\label{sec:NN-onebody}

A single particle orbital describes the motion of a single electron in an external potential. Mathematically, it is represented by a function $\phi(\bm r)$ from real space coordinates $\mathbb{R}^d$ to the space of complex numbers $\mathbb{C}$ describing amplitude and phase of the electronic wave at the given position $\bm r \in \mathbb{R}^d$. Due to the universal approximation theorem for deep feed forward neural networks \cite{Hornik1989Jan}, any such function can be approximated to arbitrary accuracy when sufficiently many hidden layers are included. Therefore, deep feed-forward neural networks are the main structural element that we employ to construct general single-particle orbitals.

The structure of the compositional wavefunction ansatz consists of three sections with distinct functionality. 
First, the function input -- i.e. the electron coordinate $\bm r$ in coordinate space in our case -- is transformed into a representative \emph{feature} 
\begin{equation}
    \bm f^0 = {\rm feature}(\bm r). 
    \label{eq:NN-orbital-feature}
\end{equation}
For systems with open boundary conditions, it is sufficient for the function ${\rm feature}(\bm r)$ to represent $\bm r$ in dimensionless units. 
Systems with periodic boundary conditions satisfy
\begin{equation}
    \Psi(\bm r_1, ..., \bm r_i, ..., \bm r_N) = \Psi( \bm r_1 ..., \bm r_i + \bm L, ..., \bm r_N)
\end{equation}
for any particle $i$, where the two vectors $\bm L = n \bm L_1 + m \bm L_2$ with $n,m \in \mathbb{Z}$ specify the supercell size and geometry.
For periodic systems, we choose the function ${\rm feature}(\bm r)$ to express the coordinate $\bm r$ in terms of periodic coordinates \cite{pescia2022neural,Cassella2023Jan},
\begin{equation}
    {\rm feature}(\bm r) = \begin{pmatrix}
        \sin(\bm G_1^T\bm r) \\
        \sin(\bm G_2^T \bm r) \\
        \cos (\bm G_1^T \bm r)\\
        \cos (\bm G_2^T \bm r)
    \end{pmatrix}
\end{equation}
where $\bm G_i^T \bm L_j = 2\pi \delta_{ij}$ are the reciprocal supercell vectors and $\delta_{ij}$ is the Kronecker delta. 
This periodic feature representation uniquely specifies the electron coordinate on the torus and ensures that proximate positions on the torus are passed as proximate features to the network, enabling efficient representation of general periodic functions by the neural network. This modification adapts the quantum chemistry implementation Psiformer of Ref.~\cite{vonGlehn2022Nov} to periodic solids. 

Second, the featured coordinate is embedded in the internal representation of the neural network by a linear transformation
\begin{equation}
    \bm h^0 = W^0 \bm f^0.
\end{equation} 
where $W_0 \in \mathbb{R}^{d_{\rm L}} \times \mathbb{R}^{2 d_{\rm dim}}$ is a rectangular matrix with $d_{\rm L}$ the width of the internal network layers. 
This embedded featured is passed through the 
multilayer perceptron neural network with functional form
\begin{equation} 
\bm h^{l + 1} = \bm h^l +  \tanh (W^{l+1} \bm h^l + \bm b^{l+1}) 
\label{eq:NN-orbital-MLP}
\end{equation}
where $W^{l+1} \in \mathbb{R}^{d_{\rm L}} \times \mathbb{R}^{d_{\rm L}}$ is the linear transformation between layers, $\bm b^{l+1} \in \mathbb{R}^{d_{\rm L}}$ is an activation bias vector, and $\tanh$ is the non-linear activation function of the nodes, and $l = 0, 1, ..., L - 1$ enumerates the layers. 

Finally, the wavefunction is obtained by projecting the output of the neural network $\bm h_{L}$ onto real and imaginary part of a complex number,
\begin{equation}
    \phi(\bm r) = \bm w_0 \cdot \bm h^{L} + i \bm w_1 \cdot \bm h^{L}
    \label{eq:NN-orbital-phi}
\end{equation} 
where $\bm w_0, \bm w_1 \in \mathbb{R}^{d_{\rm L}}$ are learnable projection vectors and ``$\cdot$'' denotes the scalar product.

\subsection{SlaterNet: Neural network for unrestricted Hartree-Fock}
\label{sec:NN-SlaterNet}

The ability to construct general single-particle wavefunctions directly enables the construction of general Slater determinant wavefunctions for $N$ electrons. This is achieved by passing each electron coordinate $\bm r_i$ through the feed-forward neural network,
\begin{align}
    \bm f^0_i & = {\rm feature}(\bm r_i) \\
    \bm h^0_i & = W^0 \bm f^0_i \\
    \bm h^{l+1}_i & = \bm h^l_i + \tanh(W^{l+1} \bm h^{l}_i + \bm b^{l+1})\ \label{eq:NN-Slater-hl}
\end{align} 
where the same transformations $W^l$ and shifts $\bm b^l$ are applied to the \emph{single-particle stream} $\bm h_i^l$ generated by the coordinates of each particle.
Thereby, this compositional function generates a single function in a high-dimensional space before projecting onto the space of $N$ single-particle orbitals $\mathbb{C}^{N}$. 
Then, $N$ distinct single-particle orbitals are constructed as 
\begin{equation}
    \phi_j(\bm r_i) = \bm w_{2j} \cdot \bm h^{L}_i + i \bm w_{2j+1} \cdot \bm h^{L}_i
    \label{eq:NN-Slater-phi}
\end{equation} 
with individually learnable projections vectors  $\bm w_{2j}, \bm w_{2j+1}$.
The projection vectors are not required to satisfy an orthogonality criterion, because non-orthogonal directions are projected out upon forming the Slater determinant as in Eq.~\eqref{eq:HF-SlaterDet}.

Because all single-electron wavefunctions are universally approximated by the deep neural network, this ansatz universally approximates all single Slater determinant wavefunctions. By minimizing the energy over this variational space, the best set of single-particle orbitals is identified. 
This is equivalent to an unrestricted Hartree-Fock calculation.
We refer to this neural network structure as "SlaterNet" throughout the manuscript.

\subsection{Self-attention neural network for electron correlations}
\label{sec:NN-attn}

Correlations due to interactions among particles can be described by specifying how the state of individual particles $i$ is modified by interactions with all remaining particles $j \neq i$. 
To capture electron correlation effects, it is necessary to go beyond Hartree-Fock ansatz as described earlier. One approach is to promote the single-particle orbitals $\phi_j(\bm r_i)$ to the \emph{correlated orbitals} $\phi_j(\bm r_i; \{ \bm r_{/i}\})$ with explicit dependence on the other particles position $\{ \bm r_{/i}\}$. The idea of correlated orbitals dates back to the backflow transformation used in variational study of superfluid helium and uniform electron gas \cite{Feynman1956Jun, Kwon1993Oct}, and has recently been applied to neural network wavefunctions \cite{Luo2019Jun,Pfau2020Sep,vonGlehn2022Nov,Li2022Dec}.

In this work, inspired by \cite{vonGlehn2022Nov}, we use the {\it self-attention} mechanism \cite{transformer} to learn how particles influence each other and how such influence affects their individual orbitals.
Specifically, the self-attention ${\rm \textsc{SelfAttn}}_i$ operates on the entire set of outputs $\{ \bm h_i^l \}$ from the previous layer and generates the set of intermediate states $\{ \bm f_i^l \}$ that is passed as input to the next perceptron layer that generates the next state $\{ \bm h_i^{l + 1} \}$,
$$ ... \to \{ \bm h_i^l \} \overset{{\rm \textsc{SelfAttn}}}{\rightarrow} \{ \bm f_i^l \} \overset{{\rm \textsc{Perceptron}}}{\rightarrow} \{ \bm h_i^{l+1} \} \to ...\ .$$
The following explains the self-attention mechanism and the mathematical expression of the compositional function is summarized in Eqs.~\eqref{eq:NN-attn-def-f} and \eqref{eq:NN-attn-def-h} below.

The first step of the self-attention operation is to define three distinct features for each element of the set $\{ \bm h_i^l \}$.  
The three features---called ``keys'', ``queries'', and ``values''---are vectors obtained by the linear transformations 
\begin{align}
    \bm k_i^{lh} & = W_{\rm k}^{lh} \bm h_i^l,\quad  W_{\rm k}^{lh} : \mathbb{R}^{d_{\rm L}} \to \mathbb{R}^{d_{\rm Attn}}\\ 
    \bm q_i^{lh} & = W_{\rm q}^{lh} \bm h_i^l,\quad  W_{\rm q}^{lh} : \mathbb{R}^{d_{\rm L}} \to \mathbb{R}^{d_{\rm Attn}} \\ 
    \bm v_i^{lh} & = W_{\rm v}^{lh} \bm h_i^l,\quad  W_{\rm v}^{lh} : \mathbb{R}^{d_{\rm L}} \to \mathbb{R}^{d_{\rm AttnValues}}
\end{align}
where the transformations $W_{\rm k}^{lh}$, $W_{\rm q}^{lh}$, and $W_{\rm v}^{lh}$ are learned and are independent of the particle index $i$.
The vector space dimensions $d_{\rm Attn}$ and $d_{\rm AttnValues}$ are typically much smaller than the dimension $d_{\rm L}$ of $\bm h_i^l$ because they only represent individual features of the state $\bm h_i^l$. 
Here, we have used multihead attention mechanism, with the index $h$ labeling independent applications of 
distinct transformations $W_{\rm k}^{lh}, W_{\rm q}^{lh}, W_{\rm v}^{lh}$ to extract multiple sets of keys, queries and values.   

The features ``key'' $\bm k_i^{lh}$ and ``query'' $\bm q_j^{lh}$ are elements of the same vector space $\mathbb{R}^{d_{\rm Attn}}$ to allow direct comparison between different electron streams $i$ and $j$.  
Specifically, in the attention mechanism the relevance of streams $j$ for a selected ``key'' streams $i$ is quantified by the similarity measure $\exp (\bm k_i^{lh} \cdot \bm q_j^{lh})$. 
By weighting the scalar product $\bm k_i^{lh} \cdot \bm q_j^{lh}$ with the exponential function, the most relevant streams $j$ for each ``key'' stream $i$ are singled out.
This procedure can be interpret as approximating the adjacency matrix of a graph describing the most relevant relations \cite{Buehler2025Jan}. 
The ``value'' feature $\bm v_j^{lh}$ quantifies the influence the stream $j$ can exert on other streams $i \neq j$.

Altogether, the self-attention operation identifies for each key stream $i$ the most relevant stream $j$ according to the feature representations $W_{\rm k}^{lh}$, $W_{\rm q}^{lh}$ and returns the value $\bm v_j^{lh}$ corresponding to the most relevant stream $j$, up to exponentially suppressed contributions from less relevant streams, 
\begin{align}
    {\rm \textsc{SelfAttn}}_i &  \left(\{ \bm h_j^l\}; W_{\rm k}^{lh}, W_{\rm q}^{lh}, W_{\rm v}^{lh}\right) \nonumber \\
     & = \frac{1}{\cal N} \sum_{j = 1}^N \exp({\bm q_j^{lh} \cdot \bm k_i^{lh}}) \bm v_j^{lh} 
     \label{eq:NN-attn-def-selfattn}
\end{align}
where a normalization factor is included
\begin{equation*}
    {\cal N} = \sqrt{d_{\rm AttnValues}} \sum_{j = 1}^N \exp({\bm q_j^{lh} \cdot \bm k_i^{lh}})\,.
\end{equation*}

The output of all attention heads $h$ is then accumulated and used to generate the input for the next perceptron layer: 
\begin{align}
    \bm f_i^{l+1} & = \bm h_i^l + W_{\rm o}^l {\rm {concat}}_h \big[ \nonumber \\ 
    & \quad {\rm \textsc{SelfAttn}}_i\left( \{ \bm h_j^l \}; W_{\rm k}^{lh},W_{\rm q}^{lh}, W_{\rm v}^{lh} \right)
 \big] \label{eq:NN-attn-def-f} \\
 \bm h^{l + 1}_i & = \bm f^{l+1}_i +  \tanh (W^{l+1} \bm f^{l+1}_i + \bm b^{l+1}) \label{eq:NN-attn-def-h}
\end{align}
where the accumulation is performed by the learned transformation $W_{\rm o}^l \in \mathbb{R}^{d_{\rm L}} \times \mathbb{R}^{N_{\rm heads} d_{\rm AttnValues}}$ projecting the output of the attention heads onto the dimension of $\bm h_i^l$. 
Equations~\eqref{eq:NN-attn-def-f} and \eqref{eq:NN-attn-def-h} replace the compositional relation Eq.~\eqref{eq:NN-Slater-hl} in the construction of one-body orbitals as in SlaterNet. 
Optionally, multiple perceptron layers can be applied between the attention layers, which is realized by multiple subsequent applications of \eqref{eq:NN-attn-def-h} with individually learnable parameters.
By mixing different single-particle streams with the self-attention mechanism, correlations between the particle states are captured.

The generalized single-particle orbitals $\phi_j(\bm r_i, \{ \bm r_{/i}\})$ constructed by the self-attention operation are permutation equivariant in the coordinates of the remaining electrons $\{ \bm r_{/i}\}$,
\begin{equation}
    \phi_j(\bm r_i;  \{ ... \bm r_l, ..., \bm r_k, ... \}) = \phi_j(\bm r_i; \{ ... \bm r_k, ..., \bm r_l, ... \})
\end{equation}
for any $k,l \neq i$. The permutation equivariance is necessary for the generalized Slater determinant to represent a fermionic wavefunction that is antisymmetric under permutation of particle coordinates. 
This property is inherent to the functional form of the self-attention operation Eq.~\eqref{eq:NN-attn-def-selfattn}.

To further enhance the expressive power of our neural network wavefunction, it is useful to construct multiple Slater determinants from the neural network output by projection onto $m = 1, ..., N_{\rm det}$ distinct sets of correlated 
orbitals for each determinant,
\begin{equation}
    \phi_j^m(\bm r_i; \{\bm r_{/i}\}) = \bm w_{2j}^m \cdot \bm h^{L}_i + i \bm w_{2j+1}^m \cdot \bm h^{L}_i
    \label{eq:NN-attn-phi}
\end{equation} 
where the projection vectors $\bm w_{2j}^m, \bm w_{2j+1}^m$ 
are individually learnable.

In summary, the full wavefunction ansatz is of the form
\begin{equation}
    \Psi(\bm R) = \sum_{m = 1}^{N_{\rm det}} \det_{ij} \left( \phi_j^m(\bm r_i; \{\bm r_{/i}\}) \right)
    \label{eq:NN-attn-final-ansatz}
\end{equation}
where the determinant is taken over the argument matrix with indices $i, j$, correlated orbitals $\phi_j^m(\bm r_i; \{\bm r_{/i}\})$ are constructed from Eq.~\eqref{eq:NN-attn-phi}.
In practice, we find using a few generalized Slater determinants usually performs better than using a single generalized Slater determinant.

\section{Variational Monte Carlo}
\label{sec:VMC}

\begin{figure}
    \centering
    \includegraphics[width=0.9\linewidth]{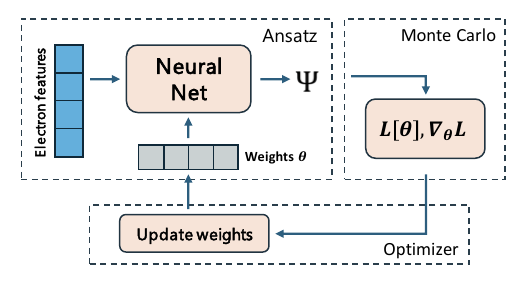}
    \caption{\textbf{Building blocks of variational Monte Carlo.} 
    In the Monte Carlo algorithm, the wavefunction ansatz $\Psi_\theta$ is constructed and sampled to  efficiently evaluate the optimization goal $L[\theta] := \langle \Psi_\theta|\hat{H}|\Psi_\theta\rangle$ of minimizing the energy. 
    Accordingly updating the weights of the variational ansatz by an optimizer during each step of the iterative procedure ensures convergence to the ground state of $\hat{H}$.
    }
    \label{fig:2}
\end{figure}

Techniques based on a variational wavefunction ansatz require efficient numerical evaluation of their energy to achieve the optimization goal of approximating the ground state of the system Hamiltonian. This is achieved by Monte Carlo sampling of the particle configuration and corresponding energy according to the distribution determined by the variational wavefunction.
This section summarizes the principal concepts of the variational Monte Carlo technique applicable to any variational wavefunction ansatz and refer to Ref.~\cite{Foulkes2001Jan} for a detailed introduction. 
A schematic overview of the Monte Carlo method is presented in Fig.~\ref{fig:2}.

\subsection{Sampling expectation values}

The optimization goal to approximate the ground state of a correlated electron Hamiltonian with a variational wavefunction ansatz $\Psi_{\bm \theta}(\bm R)$ can be chosen to minimize the energy expectation value 
\begin{equation}
    E_{\bm \theta} = \frac{\int d\bm R \Psi_{\bm \theta}^*(\bm R) \hat{H} (\bm R)\Psi_{\bm \theta}(\bm R)}{\int d\bm R |\Psi_{\bm \theta}(\bm R)|^2} 
\end{equation}
The numerically expensive integrals over configuration space $\bm R \in \mathbb{R}^{d\cdot N_{\rm el}}$, where $\bm R = (\bm r_1, ..., \bm r_N)$ are particle positions, can be efficiently evaluated by noticing the identity
\begin{align}
    E_{\bm \theta} & = \frac{\int d\bm R |\Psi_{\bm \theta}(\bm R)|^2 \Psi_{\bm \theta}^{-1}(\bm R) \hat{H} (\bm R)\Psi_{\bm \theta}(\bm R)}{\int d\bm R |\Psi_{\bm \theta}(\bm R)|^2}\ . 
\end{align}
This integral can be efficiently approximated by sampling configurations $\bm R = (\bm r_1, ..., \bm r_N)$ according to the distribution $|\Psi_{\bm \theta}(\bm R)|^2$,
\begin{align}
    E_{\bm \theta} & \approx \mathbb{E}_{\bm R \sim |\Psi_{\bm \theta}(\bm R)|^2} \left[ E_{\rm loc, {\bm \theta}} (\bm R) \right] 
    \label{eq:VMC-expectation-E_loc}
\end{align}
where the \emph{local energy} 
was introduced,
\begin{equation}
    E_{\rm loc, {\bm \theta}} = \Psi_{\bm \theta}^{-1}(\bm R) \hat{H} (\bm R)\Psi_{\bm \theta}(\bm R)
\end{equation}
and 
$\mathbb{E}_{\bm R \sim |\Psi_{\bm \theta}(\bm R)|^2} [E_{\rm loc, {\bm \theta}}] = \frac{1}{M} \sum_M E_{\rm loc, {\bm \theta}}$ denotes a summation of the local energy evaluated for $M$ sampled configurations $\bm R$ from the distribution $|\Psi_{\bm \theta}(\bm R)|^2 $.

The variance of the local energy equals the energy uncertainty $\Delta E$ of the system Hamiltonian,
\begin{align}
    (\Delta E)^2 & = \langle \hat H^2\rangle - \langle \hat H \rangle^2 \nonumber \\ 
    & \approx \mathbb{E}_{\bm R \sim |\Psi_{\bm \theta} (\bm R)|^2} \left[ E_{\rm loc, \bm\ \bm \theta}^2 (\bm R)  \right] - E_{\bm \theta}^2
\end{align}
where the equality is established by inserting $1 = \Psi_{\bm \theta}(\bm R)\Psi_{\bm \theta}^{-1}(\bm R)$ before and in between the Hamiltonian operators when computing $\langle \hat H^2\rangle$ using similar manipulations as the ones leading to Eq.~\eqref{eq:VMC-expectation-E_loc}. The energy uncertainty is proprotional to the distance $|\langle \Psi_{\bm \theta}|\Psi_0 \rangle |^2$ of the variational state $\Psi_{\bm \theta}$ from an eigenstate $\Psi_0$ of the system Hamiltonian \cite{Anandan1990Oct}. In an eigenstate, the energy uncertainty is zero.

The variational wavefunction ansatz $\Psi_{\bm \theta}$ generally is unnormalized because computing the norm itself is numerically demanding. Unnormalized probability distributions can be sampled using the Metropolis-Hastings algorithm \cite{Metropolis1949Sep,Hastings1970Apr}; this technique is applied in the numerical results presented in this manuscript. 

\subsection{Optimizing variational parameters}

To optimize the parameters of the wavefunction, the steepest descent direction is identified as the direction that minimizes the cost function $L[\bm \theta + d\bm \theta]$ for a step $d\bm \theta$, where the distance of the step is measured on the variational wavefunction space $|| d\bm \theta ||_\Psi$ \cite{Amari1998Feb,Sorella1998May},
\begin{align}
    || d\bm \theta ||_\Psi^2 = 1 - |\langle \Psi_{\bm \theta  + d\bm \theta} | \Psi_{\bm \theta}\rangle|^2 \approx \sum_{n,m} g_{nm}(\bm \theta) d\theta_n d\theta_m 
\end{align}
up to higher order corrections, where the summation includes all variational parameters $\theta_n$ and $g_{nm}(\bm \theta)$ is the quantum geometric tensor
\begin{equation}
    g_{nm}(\bm \theta) = \langle \partial_{\theta_n} \Psi_{\bm \theta } | \left(1 - |\Psi_{\bm \theta}\rangle\langle\Psi_{\bm \theta}| \right)| \partial_{\theta_m} \Psi_{\bm \theta }\rangle \ .
    \label{eq:VMC-geometric-tensor}
\end{equation}
In this way, the information of how each parameter affects the wavefunction is included in the distance measure. The resulting optimal step then includes the inverse of the quantum geometric tensor \cite{Amari1998Feb},
\begin{equation}
    d\bm \theta = -\eta g^{-1}(\bm \theta) \nabla_{\bm \theta} L[\bm \theta] \ .
    \label{eq:VMC-steepest-descent}
\end{equation}
where $\eta$ controls the length of the optimizer step $d \bm \theta$ and is commonly known as the \emph{learning rate}. 
This procedure is known as natural gradient descent. 

In practice, for variational models with large number of parameters, computing the inverse of the quantum geometric tensor is numerically impractical. For neural network wavefunction ans\"atze, the inverse of the geometric tensor can be efficiently approximated by the Kronecker-factorized approximate curvature (KFAC) method  \cite{Martens2015Jun,Pfau2020Sep}, where correlations between different layers as well as input and output of individual layers are neglected. Following Ref.~\cite{Pfau2020Sep}, this method is applied in our numerical calculations. The current implementation of KFAC operates only on the absolute magnitude of the wavefunction, while the phase is neglected. In this case, the quantum geometric tensor reduces to the Fisher information matrix.

We remark that around a second-order quantum phase transition, the Fisher information diverges in the parameter that drives the quantum phase transition \cite{CamposVenuti2007Aug,Gu2008Nov}. These divergencies may pose difficulties in the numerical approximation of the Fisher information matrix so that optimization around a quantum critical point may be more challenging than deep within a phase -- even if the variational ansatz remains sufficiently expressive to describe the critical state.

\section{Results}
\label{sec:results}

This section presents the numerical results obtained from our variational Monte Carlo calculations of Hartree-Fock using SlaterNet and the minimal self-attention NN and the benchmark with band-projected exact diagonalization. 
 
Throughout this article, we consider ${\rm WSe}_2 / {\rm WS}_2$ as model system, with the following model parameters determined by first-principles calculations \cite{mott2}: effective mass $m^* = 0.35 m_e$, moir\'e potential strength $V_0 = 15\ {\rm meV}$, and moir\'e shape parameter $\varphi = \pi / 4$. We consider a moir\'e lattice period of $a_M = 8.031 \ {\rm nm}$ which results from the lattice mismatch between WSe$_2$ and WS$_2$. 
In this case, moir\'e filling of $\nu = 1$ particles per moir\'e unit cell corresponds to a density of $n = 1.785 \times 10^{12} \ {\rm cm}^{-2}$. The relative dielectric constant for a surrounding dielectric hBN is $\epsilon \approx 5$. Devices with tunable dielectric constants are possible using tunable dielectrics such as ${\rm Sr}{\rm Ti}{\rm O}_3$ \cite{Fuchs1999May}. All our calculations are performed for the spin-polarized system. 

These units are converted to the dimensionless units of Eq.~\eqref{eq:system-hamiltonian} as follows. In Eq.~\eqref{eq:system-hamiltonian}, distances are measured in effective Bohr radii $a^*_{B} = 4\pi\epsilon\epsilon_{0}\frac{\hbar^{2}}{m^{*}e^{2}} = \epsilon \frac{m_e}{m_*} a_0$ and energies in effective Hartree $\rm{Ha}^* = \frac{\hbar^{2}}{m^{*}{a_0^*}^{2}} = \frac{1}{\epsilon^2}\frac{m^*}{m_e} \rm{Ha}$, where $a_0$ and ${\rm Ha}$ are the standard Bohr radius and Hartree energy defined in terms of the free electron mass $m_e$, $m^*$ is the effective mass for charge carriers, and $\epsilon$ is the relative dielectric constant of the surrounding medium. 

We summarize the hyperparameters used for training the models in Table \ref{Tab:Hyperparams} in the App.~\ref{app:Hyperparameters}. While architectural parameters must be adjusted for each specific system (as detailed below), we found that the remaining hyperparameters were largely independent of the problem's specifics and consistently yielded stable and optimal results. We particularly highlight the learning rate $\eta_0 = 10$, which, although unusually high, emerged as the optimal value for a wide variety of 2D systems we experimented with, including both free Fermi liquid and Moiré systems.

\subsection{Convergence and scaling with system size}

\begin{figure}
    \centering
    \includegraphics[width=\linewidth]{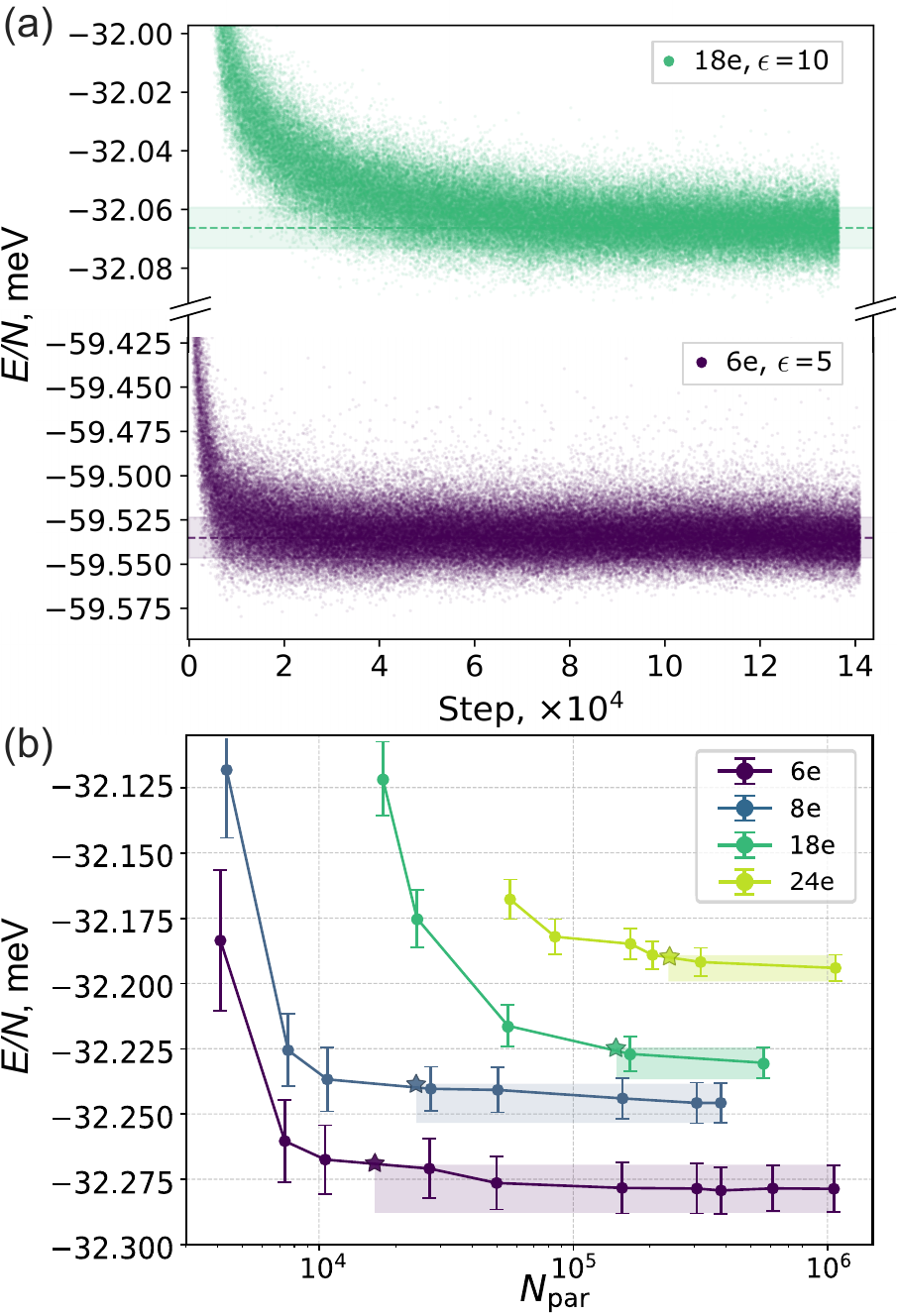}
    \caption{(a) Raw (unsmoothed) learning curves for the energy per electron at $\nu=2/3$ filling for 6 electrons with $\epsilon = 5$, and 18 electrons with $\epsilon = 10$.
    (b)
    Converged energy of the moir\'e system at $\nu = 2/3$ filling and $\epsilon = 10$ as a function of the number of variational parameters in the self-attention NN. Error bars indicate the standard deviation of the mean local energy, averaged over the batch size at the end of optimization. Stars mark the estimated number of parameters, $N^{*}_\text{par}$, required to reach convergence saturation. Further increasing in the number of parameters keeps the energies within one standard deviation (highlighted by the striped region). For sake of presentation, the $8e$ data was shifted up by $0.39$meV, and $18e$ and $24e$ datas shifted down by $0.16$meV and $0.25$meV, respectively. Default simulation parameters are contained in Table~\ref{Tab:Hyperparams}. } 
    \label{fig:3}
\end{figure}

In any numerical method, it is crucial to ensure convergence and address how the numerical complexity scales with the system size. 
Here, we first demonstrate convergence of the energy estimate as a function of Monte Carlo steps. 
Next, we investigate the saturation of the converged energy with the network dimensions.
This enables us to numerically estimate the numerical complexity in terms of the scaling of the required number of variational parameters with the number of electrons.

To ensure the convergence of our results, we verified that the learning curves ({\it i.e.} the local energy averaged of the batch size of each step) exhibited stable behavior and, on average, decreased monotonically. This trend remained consistent across different system sizes, as illustrated by the raw learning curves for the 9- and 27-site systems at $\nu=2/3$ filling in Fig. \ref{fig:3}(a), shown without smoothing. We highlight the mean and standard deviation of the obtained mean-energy-per-batch using a dashed line and a stripe. The latter arises from the finite batch size used for sampling at each iteration and sets an important scale, as independent learning curves typically remain within one standard deviation of the mean.

To further validate our approach, we examine the convergence of the ground-state energy per particle as a function of the number of variational parameters for the 9-, 12-, 27- and 36-site systems at 
$\nu=2/3$ filling. Fig. \ref{fig:3}(b) explores networks spanning three orders of magnitude in variational parameters and reveals a clear convergence pattern in all cases, with the ground-state energy approaching a lower bound as the network reaches 1 million variational parameters.

This distinct converging pattern indicates \textit{convergence saturation}, meaning that further increasing the number of parameters does not improve the ground-state energy. We define the saturation point 
$N^*_{\text{par}}$ for a given system as the threshold beyond which the converged energies consistently fall within one standard deviation of the lowest observed energy. In Fig. \ref{fig:3}(b), these values are marked with stars.

\begin{figure}
    \centering
    \includegraphics[width=0.9\linewidth]{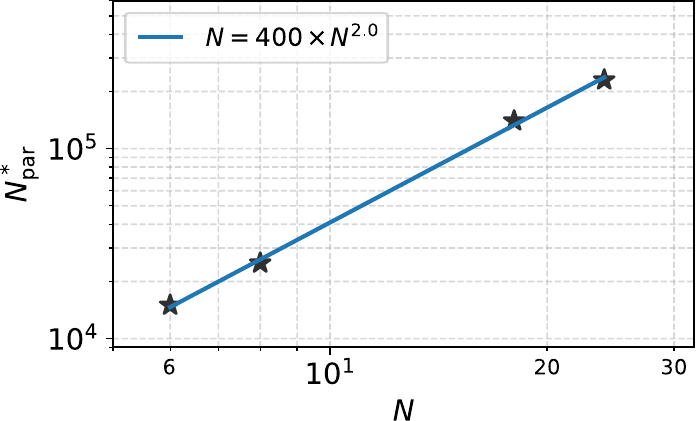}
    \caption{Scaling law for the number of variational parameters at convergence saturation as a function of the number of electrons for $\epsilon = 10$. The extracted saturation points are well described by a power-law relation $N^* = C \cdot N^\alpha$, with best-fit parameters $\alpha = 2.01 \pm 0.05$ and $C = (4.00 \pm 0.49) \times 10^2$. This relation quantifies how the required parameter count grows with system size and provides a practical guideline for selecting an optimal network size while ensuring convergence.
 }
    \label{fig:4}
\end{figure}

We extract these saturation points as a function of the number of electrons $N$ and find that they follow a well-defined scaling law:
\begin{equation} 
N_\text{par}^* \approx (400 \pm 49) \times N^{2.01 \pm 0.05} \ .
\end{equation}
The fit is shown in Fig.~\ref{fig:4}.
While this scaling behavior is derived for the present system and network architecture, we believe it may serve as a useful approximation for estimating parameter growth in other settings as well, though it may not directly generalize. 

This scaling law provides a helpful estimate for the number of parameters required to achieve a given accuracy; however, the parameter count alone is not the only relevant factor. Since attention heads play a crucial role in capturing correlations, a minimum number of attention heads and layers (approximately above 3 in our tests) is necessary to reach the ground state. Simply increasing the total number of parameters without ensuring sufficient attention heads and layers would not improve performance. Our scaling law serves as a guideline within the regime where these minimal requirements are met. We leave a more detailed analysis of scaling laws as a function of individual network dimensions for future work.

In comparison, tensor network methods for two-dimensional lattices under area-law assumptions require a number of parameters scaling as $\propto e^{\sqrt{N}}$ \cite{Ganahl2023Feb}. For tensor networks, parameter scaling depends on the entanglement structure of the problem and tensor network ansatz. Our empirical result of quadratic scaling of the parameters in the self-attention NN ansatz suggests significantly better scaling compared to tensor network methods.

\subsection{Benchmark for small system size}

After ensuring convergence and saturation in the number of variational parameters, we benchmark our \FNN results with SlaterNet and BP-ED to assess the quantitative accuracy of predictions from the self-attention NN. 

\begin{figure}
    \centering
    \includegraphics[width=\linewidth]{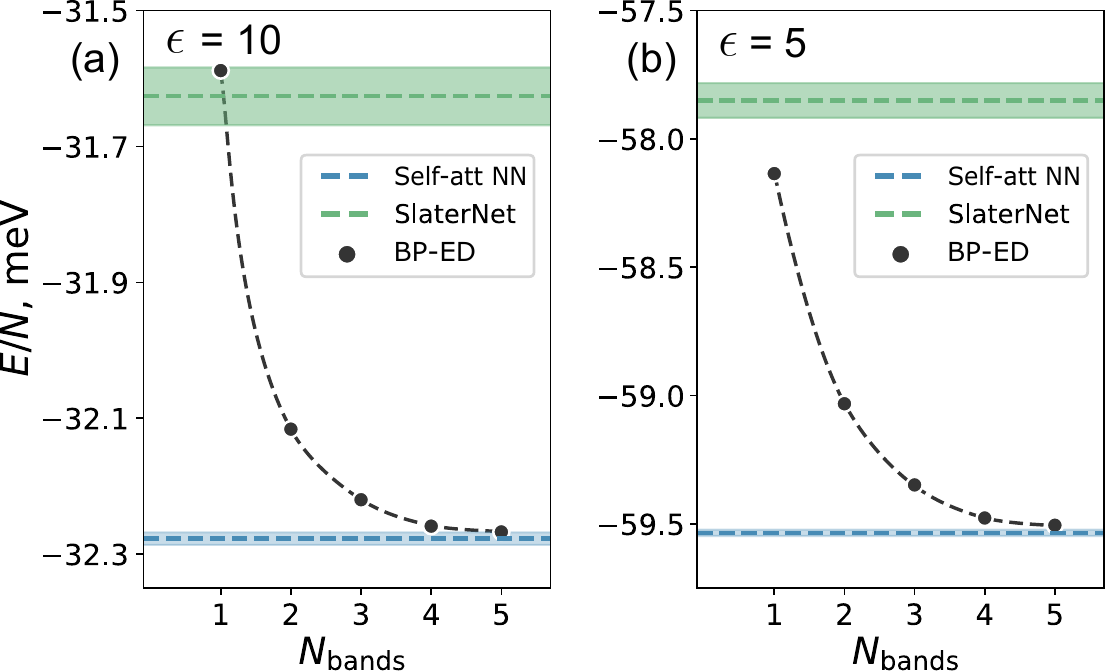}
    \caption{Comparison of ground state energies obtained from the self-attention NN (blue), Hartree-Fock from SlaterNet (green), and BP-ED (dots) in a $3\times3$ supercell with $\nu = 2/3$ filling. Dashed lines for the self-attention NN and SlaterNet indicate mean of local energy obtained from the variational Monte Carlo algorithm. 
    The shaded region indicates the standard deviation of the mean of the local energy averaged over the batch size per optimization step. Dashed gray line is a guide to the eyes.
    }
    \label{fig:5}
\end{figure}

We study a variety of different system sizes and regimes at different interaction strengths. The comprehensive energy comparisons given in Fig.~\ref{fig:5} graphically evince the difference in ground state energy levels between all 3 methods for 9 unit cell system again at $\nu=2/3$. The \FNN energy is considerably lower than SlaterNet, with the difference precisely representing the correlation energy. We see that the correlation energy is about $2\%$ in our moiré system, considerably higher than the correlation energy in ordinary molecular systems \cite{Foulkes2001Jan}. 
This enhanced correlation energy underscores the relevance of moiré systems for studying strongly correlated phenomena. Additionally, we observe that as the number of bands increases in BP-ED, the ground-state energy approaches but remains higher than that of the \FNN.

The advantage of \FNN becomes even more pronounced in larger systems, such as those with 18 electrons [Table~\ref{Tab:Comparison_18e}], where BP-ED is limited to a single band due to the enormous Hilbert space dimension. In this case, the \FNN energy is nearly $2.5\%$ lower than that of BP-ED. In fact, the band mixing is so strong that even SlaterNet achieves a lower energy than BP-ED.

These results demonstrate that \FNN consistently achieves lower energies than both BP-ED and SlaterNet, underscoring its capacity to express ground states of the moir\'e systems to higher accuracy, and revealing the variational superiority of our \FNN method.

\begin{table}[h]
    \centering
    \begin{tabular}{ccccc}
        \toprule
        $N_e$ & $\epsilon$ & \quad self-attention NN & \quad SlaterNet & \quad BP-ED\\
        \midrule
        18 & 10 &\quad \textbf{-32.070(7)} &\quad -31.35(2) &\quad -31.32443\\
        \midrule
        18 & 5 &\quad \textbf{-59.127(9)} &\quad -58.01(3) &\quad -57.80848\\
        \bottomrule
    \end{tabular}
    \caption{Comparison of ground-state energies of the 27 site system at $2/3$ filling and $\epsilon = 10$ and $5$ obtained using different methods. For 27 sites, BP-ED is limited to a single band due to the prohibitively large Hilbert space dimension.}
    \label{Tab:Comparison_18e}
\end{table}

\subsection{Fermi Liquids and Generalized Wigner Crystals}

With the variational superiority of the \FNN method established, we now demonstrate the capability of the \FNN to not only reveal the essential physics of the strongly correlated system, but predict expected phenomenon. In real homobilayer systems, experimentalists can adjust displacement field to tune the level of electron-electron interactions, driving the system from a Fermi liquid state to generalized Wigner crystal state as interactions increase. While heterobilayer systems do not offer the same experimental freedom, a metal-insulator transition have been theoretically predicted as electron interactions grow stronger by previous ED and QMC studies \cite{morales2021metal, yang2024metal}. \par

\begin{figure}
    \centering
    \includegraphics[width=0.9\linewidth]{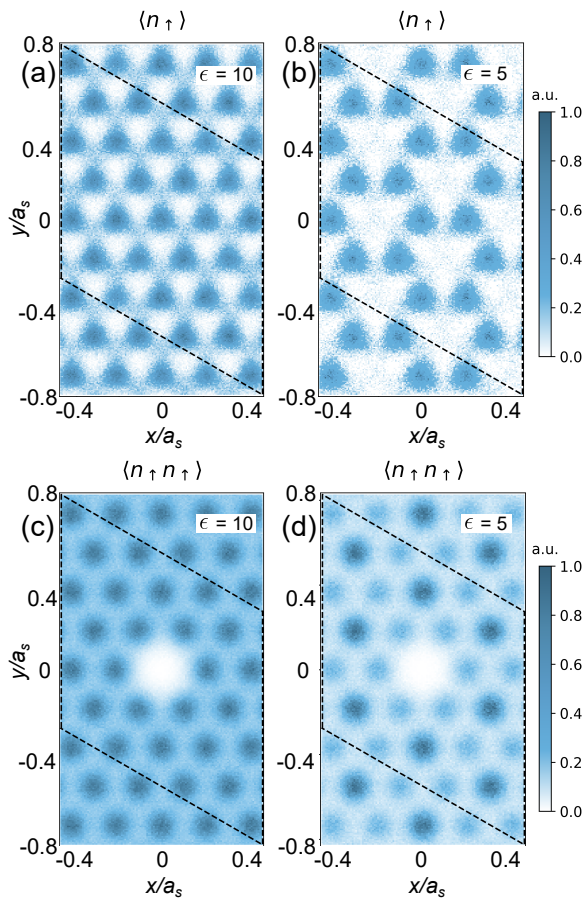}
    \caption{Converged ground state charge density [(a), (b)] and density-density correlation function [(c), (d)] the moir\'e system at $\nu = 2/3$ filling of 27 sites per supercell with $\epsilon = 10$ (left) and $\epsilon = 5$ (right column).
    The dashed line indicates the boundary of the periodic simulation supercell.}
    \label{fig:6}
\end{figure}

Numerically, we can simulate this phenomena by changing the dielectric constant $\epsilon$ which is inversely proportional to interaction strength. When interactions are small compared to kinetic energy, charge carriers are de-localized and the system is in the free-electron regime. In the presence of weak interactions, electrons will interact with each other at the boundary of the Brillouin zone, giving rise to a Fermi-liquid. In real space, the delocalization of electrons results occupation of all potential minimum sites of the superlattice, producing the triangular lattice density pattern shown in Fig. \ref{fig:6} a). 

As interactions increase, the Coulomb repulsion between the charge carriers make it more energetically favorable to spread as far apart as possible, driving the system to transition into an electron crystalline state. The crystal breaks the translational symmetry of the underlying potential. At $\nu = 2/3$ electrons per triangular potential unit cell, the electronic crystal has the shape of a honeycomb lattice as shown by the density pattern in Fig.~\ref{fig:6} b).

In addition to the density profile, Fig.~\ref{fig:6} (c) and (d) show the pair correlation function for weak ($\epsilon=10$) and strong interaction ($\epsilon=5$), respectively. 
Both the density and pair correlation profile show clear signatures of a Fermi liquid state for $\epsilon=10$ and a generalized Wigner crystal state for $\epsilon=5$. 
This demonstrates the \FNN's ability to capture the essential physics and strongly correlated behavior beyond simply lowering the ground state energy. 

Finally, the difference in energies from the \FNN and SlaterNet allows to estimate correlation energy Eq.~\eqref{eq:methods-correlation-energy}. We include the extracted correlation energy as a function of dielectric constant in App.~\ref{app:Correlation}. 
 Furthermore, in App.~\ref{app:ED-benchmark}, we provide additional exact-diagonalization results on the metal-insulator phase transition.

\section{Comparison to other neural network ans\"atze}

In contrast to previous works studying periodic solids with neural quantum states \cite{Cassella2023Jan,Li2022Dec}, we do not multiply an envelope function when constructing single-particle orbitals. Instead the full functional dependence of the periodic orbitals is represented by the neural network. Ref.~\cite{Cassella2023Jan} motivated the use of envelope functions by a better convergence for their system; however our numerical experiments for our system did not show any improvements when envelope functions are included. Ref.~\cite{Li2022Dec} found faster convergence using plane wave envelopes when the system was close to a Hartree-Fock state. Envelope functions introduce human bias towards a certain functional dependence, therefore we remove them in our network architecture. 

Further, in our wavefunction ansatz, we do not multiply a Jastrow factor to the wavefunction constructed from generalized Slater determinants; thereby deviating from the PsiFormer architecture for molecular systems \cite{vonGlehn2022Nov}. A Jastrow factor is usually motivated to capture correctly the wavefunction cusps that occur when two particles approach each other, which is model specific and can be derived analytically, see App.~\ref{app:Jastrow} and Ref.~\cite{Foulkes2001Jan} for details. 
We performed numerical experiments [App.~\ref{app:Jastrow}] including a simple Jastrow factor with a single learnable parameter that only enforces the correct wavefunction shape as two particles approach each other. We did not find significant improvement by including the Jastrow factor, instead we observed that including the Jastrow increased the GPU time per step by around 10 to 20$\%$. 
In contrast to molecular systems, we suspect that the Jastrow factor is less relevant for low-density electrons in a periodic potential as studied in this manuscript, because electrons are mostly bound to individual wells while overlap of two electrons is small. 

We highlight that our results have been computed from a randomly initialized wavefunction. 
The only physical input is the determinant operation guaranteeing the antisymmetric structure of the fermionic wavefunction. 
In contrast, previous works \cite{Pfau2020Sep,vonGlehn2022Nov} applied pretraining with Hartree-Fock to initialize the wavefunction before optimizing energy. Also, recent message-passing architectures \cite{pescia2023message} describe correlations starting from a chosen Slater determinant of single-particle orbitals.

In the future, our work can be extended to include computation of excited states \cite{Pfau2024Aug} and zero-temperature Green's functions and spectral functions  \cite{Hendry2021Nov}. Finite temperature effects can be studied by an extension of the neural network architecture to represent thermal density matrices \cite{Irikura2020Mar,Lu2024Jan}. Dynamical and open systems can be studied by applying the time-dependent variational principle \cite{Jackiw1979Apr} to a variational density matrix \cite{Hartmann2019Jun,Nagy2019Jun,Vicentini2019Jun,Yoshioka2019Jun,Reh2021Dec,Mellak2024Aug}. 

\section{Conclusion}

In this work, we studied the performance of a wavefunction ansatz based on a many-parameter, self-attention neural network with minimal human input at a case study of a periodic moir\'e solid. Our numerical experiments revealed that
\begin{itemize}
    \item the self-attention neural network accurately predicts ground state energy, outperforming our benchmark with band-projected exact diagonalization 
    \item the minimal number of parameters required for energy convergence scales roughly as $N^2$ with the particle number, suggesting an efficient description of the ground state
    \item charge density and correlation functions identify a Fermi-liquid to Wigner crystal phase transition, confirming convergence on both sides of the phase transition
\end{itemize}
Furthermore, Ref.~\cite{Teng2024Nov} recently demonstrated that a similar self-attention neural network is able to accurately describe fractional quantum Hall ground state wavefunctions, demonstrating that the ansatz is able to describe ground states whose correlations are fundamentally distinct from a state of uncorrelated orbitals.

The conceptual strength of the self-attention neural network ansatz lies in its two pillars of construction: (i) it employs a large scale parameterization without human bias, (ii) the self-attention mechanism efficiently learns relations between electrons that encode their correlations. 
Altogether, these results put faith in the hope that a unified wavefunction ansatz may be able to accurately describe a wide range of, if not all, quantum phases in strongly interacting electron systems.

{\it Acknowledgements---.} We are grateful to Aidan Reddy for assistance with exact diagonalization study. We thank Tony Teng, Karen Lei, Xiang Li, and Yixiao Chen for helpful discussions. This work was primarily supported by a Simons Investigator Award from the Simons Foundation. M.G. acknowledges support from the German Research
Foundation under the Walter Benjamin program (Grant
Agreement No. 526129603) and by the Air Force Office of Scientific Research under award number FA2386-24-1-4043.
TZ was supported by the MIT Dean of Science Graduate Student Fellowship.  KN was supported by Air Force Office of Scientific Research under award number FA2386-21-1-4058. LF also acknowledges support from the NSF through Award No. PHY-2425180. 
This work made use of resources provided by subMIT at MIT Physics.
The authors acknowledge the MIT SuperCloud and Lincoln Laboratory Supercomputing Center for providing resources \cite{reuther2018interactive}.
 Our software implementation is based on Google DeepMind's VMC package FermiNet \cite{ferminet_github} and makes use of the Forward Laplacian \cite{Li2023JulForwardLaplacian} implementation from \cite{fwdlap_github}.


\begin{appendix}

\section{Coulomb system with periodic boundary condition}
\label{app:ewald}

In the following, we employ and develop numerical techniques for solving the interacting Hamiltonian Eq.~\eqref{eq:system-hamiltonian} on a finite-size cluster with periodic boundary condition, such that the $N$-electron wavefunction satisfies the condition 
\begin{equation}
    \Psi(\bm r_1, ..., \bm r_i, ..., \bm r_N) = \Psi( \bm r_1 ..., \bm r_i + \bm L, ..., \bm r_N)
\end{equation}
for any particle $i$, where the two vectors $\bm L = n \bm L_1 + m \bm L_2$ with $n,m \in \mathbb{Z}$ specify the cluster size and geometry. For systems with an underlying periodic potential as in Eq.~\eqref{eq:system-hamiltonian}, the cluster must accommodate an integer number of unit cells of the periodic potential. Our finite size system can also be viewed as one supercell within an infinitely large 2D plane in which only particle configurations satisfying supercell periodicity are allowed. 

For systems with periodic boundary conditions, the Coulomb interaction must be adapted to account for the interaction energy with periodic images of the charge configuration in the other ``supercells''. The Coulomb interaction decomposes into two terms,
\begin{equation}
    H_{ee}=\frac{1}{2}\sum_{i}^{N}\sum_{i\neq j}^{N}\sum_{\bm{L}}\frac{1}{|\bm{r}_{i}-\bm{r}_{j}+\bm{L}|}+\frac{1}{2}\sum_{i}^{N}\sum_{\bm{L}\neq0}\frac{1}{|\bm{L}|}
    \label{eq:system-ewald-Hee-periodic}
\end{equation}
where the first term describes interaction of the $N$ distinct electrons in the supercell and their periodic images, while the second term describes the interaction of each individual electron in the supercell with its own periodic images. The second terms is determined by the \emph{Madelung
constant} $\xi_{{\rm M}}=\sum_{\bm{L}\neq0}\frac{1}{|\bm{L}|}$. 
Because Coulomb interactions are long-ranged, the summation contains a divergent, homogeneous contribution. This divergence is canceled when a neutralizing charge background is introduced, as we show later.

The remainder of this section describes how the Coulomb interaction and Madelung constant are efficiently evaluated using \emph{Ewald summation}. Our derivation closely follows Ref.~\cite{Fraser1996Jan}. Readers not interested in this derivation can refer to Eqs.~\eqref{eq:system-Hee-Ewald-result} and \eqref{eq:system-Madelung-Ewald-result} for the result of the electron-electon interaction and Madelung constant, respectively, that defines the Coulomb Hamiltonian for our numerical study. 

The numerical procedure for the efficient computation of Eq.~\eqref{eq:system-ewald-Hee-periodic} proceeds by evaluating the Coulomb potential induced by an individual electron at position $\bm{r}_{a}$ in the supercell and its periodic images,
\begin{equation}
    \varphi_{a}(\bm{r})=\sum_{\bm{L}}\frac{1}{|\bm{r}-\bm{r}_{a}-\bm{L}|}
\end{equation}

The principal idea of Ewald summation is to decompose the sum into quickly converging
sums for short-ranged and long-ranged contributions. This is achieved
by Fourier transformation of each individual term, where the integration
domain is taken as the infinite plane,
\begin{equation}
    \varphi_{a,\bm{L}}(\bm{q})=\int_{\mathbb{R}^2} d{\bm r}\frac{e^{i\bm{q}\bm{r}}}{|\bm{r}-\bm{r}_{a}-\bm{L}|}=2\pi\frac{e^{-i\bm{q}(\bm{r}_{a}+\bm{L})}}{q}
\end{equation}
where $q=|\bm{q}|$. By applying the identity
\begin{equation}
    \frac{1}{q}=\frac{2}{\sqrt{\pi}}\int_{0}^{\infty}dte^{-q^{2}t^{2}}=\frac{2}{\sqrt{\pi}}\left[\int_{0}^{\eta}+\int_{\eta}^{\infty}\right]dte^{-q^{2}t^{2}}
\end{equation}
the integration is split into short-ranged and long ranged contributions.
 The first term describes the short-ranged contributions, as is seen
by Fourier transformation back to real space,

\begin{align}
\varphi_{a,\bm{L}}^{{\rm S}}(\bm{r}) & =\frac{1}{\pi^{3/2}}\int_{\mathbb{R}^{2}}d^{2}\bm{q}e^{i\bm{q}(\bm{r}-\bm{r}_{a}-\bm{L})}\int_{0}^{\eta}dte^{-q^{2}t^{2}} \nonumber \\
 & =\frac{1}{|\bm{r}-\bm{r}_{a}-\bm{L}|}\left[{\rm Erfc}\left(\frac{|\bm{r}-\bm{r}_{a}-\bm{L}|}{2\eta}\right)\right]
\end{align}
The integral describing the long-ranged contributions can be directly
evaluated in momentum space,
\begin{equation}
    \varphi_{a,\bm{L}}^{{\rm L}}(\bm{q})=2\pi e^{-i\bm{q}\bm{r}_{a}}\frac{{\rm Erfc}\left(\eta q\right)}{q}\,.
\end{equation}
The complimentary error function ${\rm Erfc}(x)$ ensures fast convergence
of the summation due to an exponential suppression for $x>1$. 

The long-range part contains a divergent contribution in the limit
$q\to0$. By requiring that the mean electrostatic potential is zero,
the $\bm{q}=0$ contributions of both short- and long-range potentials
are cancelled by a homogeneous charge background. This cancels the
divergence of the long-range part, and requires to subtract the $\bm{q}\to0$
contribution of the short-range part, 
\begin{align}
\varphi_{a,\bm{q}\to0}^{{\rm S}} & =\int_{\mathbb{R}^{2}}\frac{d^{2}\bm{q}}{2\pi}\sum_{\bm{L}}e^{i\bm{q}(\bm{r}-\bm{r}_{a}-\bm{L})}\frac{{\rm Erf}\left(\eta q\right)}{q}\nonumber \\
 & =\int_{\mathbb{R}^{2}}\frac{d^{2}\bm{q}}{2\pi}\sum_{\bm{G}}\delta(\bm{q}-\bm{G})e^{i\bm{q}(\bm{r}-\bm{r}_{a})}\frac{{\rm Erf}\left(\eta q\right)}{q}\nonumber \\
 & =\frac{2\pi}{A_{{\rm u.c.}}}\sum_{\bm{G}}e^{i\bm{G}(\bm{r}-\bm{r}_{a})}\frac{{\rm Erf}\left(\eta G\right)}{G}\nonumber \\
 & =\frac{2\pi}{A_{{\rm u.c.}}}\frac{2\eta}{\sqrt{\pi}}
\end{align}
The total electrostatic potential of a charge at $\bm{r}_{a}$ and
its periodic images is given by the summation of short- and long-range
contributions from all unit cells shifted by $\bm{L}$,
\begin{align}
    \varphi_{a}(\bm{r})&=\sum_{\bm{L}}\varphi_{a,\bm{L}}^{{\rm S}}(\bm{r})-\varphi_{a,\bm{q}\to0}^{{\rm S}}\nonumber \\ 
    & \  + \sum_{\bm{L}}\int_{\mathbb{R}_{/\bm{0}}^{2}}\frac{d^{2}\bm{q}}{4\pi^{2}}e^{i\bm{q}\bm{r}}\varphi_{a,\bm{L}}^{{\rm L}}(\bm{q})
\end{align}
The long-range contribution can further be simplified,
\begin{align}
& \ \sum_{\bm{L}}\int_{\mathbb{R}_{/\bm{0}}^{2}}\frac{d^{2}\bm{q}}{4\pi^{2}}e^{i\bm{q}\bm{r}}\varphi_{a,\bm{L}}^{{\rm L}}(\bm{q}) \nonumber \\
 & =\frac{1}{2\pi}\int_{\mathbb{R}_{/\bm{0}}^{2}}d^{2}\bm{q}\sum_{\bm{G}}\delta(\bm{q}-\bm{G})e^{i\bm{q}(\bm{r}-\bm{r}_{a})}\frac{{\rm Erfc}\left(\eta q\right)}{q} \nonumber\\
 & =\frac{2\pi}{A_{{\rm u.c.}}}\sum_{\bm{G}\neq0}e^{i\bm{G}(\bm{r}-\bm{r}_{a})}\frac{{\rm Erfc}\left(\eta G\right)}{G}
\end{align}
where $\bm{G}$ are reciprocal supercell vectors $\bm{G}\cdot\bm{L}=2\pi n$,
$n\in\mathbb{N}$. 

The total electrostatic energy of a configuration of $N$ particles
in a unit cell is then written as
\begin{equation}
    H_{ee}=\frac{1}{2}\sum_{b}\sum_{a\neq b}\varphi_{a}(\bm{r}_{b})+\frac{1}{2}\sum_{b}\xi_{{\rm M}}
    \label{eq:system-Hee-Ewald-result}
\end{equation}
where $a,b=1,...,N$ enumerates the particles in the unit cell, the
first term describes the energy of all particles $b$ in the total
potential of all other particles $a\neq b$ while the second term
describes the energy of all particles in the potential created by
their images in the other unit cells, determined by the Madelung constant,
\begin{align}
\xi_{{\rm M}} & =\sum_{\bm{L}\neq0}\varphi_{b,\bm{L}}^{{\rm S}}(\bm{r}_{b})+\sum_{\bm{L}\neq0}\int_{\mathbb{R}_{/\bm{0}}^{2}}\frac{d^{2}\bm{q}}{4\pi^{2}}e^{i\bm{q}\bm{r}_{b}}\varphi_{b,\bm{L}}^{{\rm L}}(\bm{q}) \nonumber \\
 & =\sum_{\bm{L}\neq0}\frac{1}{|\bm{L}|}\left[{\rm Erfc}\left(\frac{|\bm{L}|}{2\eta}\right)\right]-\varphi_{a,\bm{q}\to0}^{{\rm S}} \nonumber\\
 & \ +\frac{2\pi}{A_{{\rm u.c.}}}\sum_{\bm{G}\neq0}\frac{{\rm Erfc}\left(\eta G\right)}{G}-\xi_{0}^{{\rm L}} 
 \label{eq:system-Madelung-Ewald-result}
\end{align}
where the $\bm{q}\to0$ part of the short-range potential is subtracted,
and the term $\xi_{0}^{{\rm L}}$ arises from subtracting the $\bm{L}=0$
term from the evaluation of the long-range potential
\begin{align*}
\xi_{0}^{{\rm L}} =\int \frac{d^{2}\bm{q}}{2\pi}\frac{{\rm Erfc}\left(\eta q\right)}{q}=\frac{1}{\eta \sqrt{\pi}} \ .
\end{align*}

For exact diagonalization in momentum space as presented in Sec.~\ref{sec:traditional_methods}, we note that the reciprocal space Coulomb interaction [last term in Eq.~\eqref{eq:methods-ED-H}] arises from Fourier transform of the first term of Eq.~\eqref{eq:system-ewald-Hee-periodic} describing the real-space Coulomb interaction in systems with periodic boundary condition. 
In the reciprocal space formulation of Eq.~\eqref{eq:methods-ED-H}, the divergent $\bm q = 0$ term of the Coulomb interaction is set to zero by requiring cancellation with a homogeneous charge background; the same requirement was imposed when deriving the real-space Coulomb interaction [Eq.~\eqref{eq:system-Hee-Ewald-result}] using Ewald summation. 
Notice that the Madelung energy $N \xi_{\rm M}/2$ needs to be explicitly included in the reciprocal space formulation [Eq.~\eqref{eq:methods-ED-H}]] in order to capture the interaction of each electron in the supercell with its own periodic images.

\section{Jastrow factors}
\label{app:Jastrow}

A conventional approach to describe electron-electron correlations beyond Hartree-Fock is to multiply a \emph{Jastrow} factor to a Slater determinant \cite{Jastrow1955Jun}. No universal approximation theorem exists for this wavefunction ansatz to the knowledge of the authors at the time of publication. Jastrow factors are typically be model specific and physically motivated, allowing to capture a large fraction of the correlation energy \cite{Foulkes2001Jan}.

A Jastrow factor is a bosonic (i.e. permutation equivariant) function that is multiplied to the many-body wavefunction, 
\begin{equation*}
    \Psi(\bm R) \to J(\bm R) \Psi(\bm R)\ .
\end{equation*}

For systems with Coulomb interactions, a typical choice for the Jastrow factor is a functional of the form
\begin{equation}
    J(\bm R) = e^{- \sum_{i }^{N_{\rm elec}}\sum_{i \neq j}^{N_{\rm elec}} u(||\bm r_i - \bm r_j||)}
    \label{eq:NN-attn-Jastrow-def}
\end{equation}
where $u(r)$ depends on the relative electron-electron distances and is tailored to capture the wavefunction cusps as two electrons approach each other: 
For Coulomb interaction, the energy diverges as two particles approach each other. This divergence is canceled by a divergence of kinetic energy contribution for a specific functional form of the wavefunction. Because eigenstates of the Coulomb gas have finite energy, this cancellation occurs for all eigenstates, and the functional form as two particles approach each other can be analytically derived. 
This dependence can be explicitly enforced by the Jastrow factor. 

\begin{figure}
    \centering
    \includegraphics[width=0.9\linewidth]{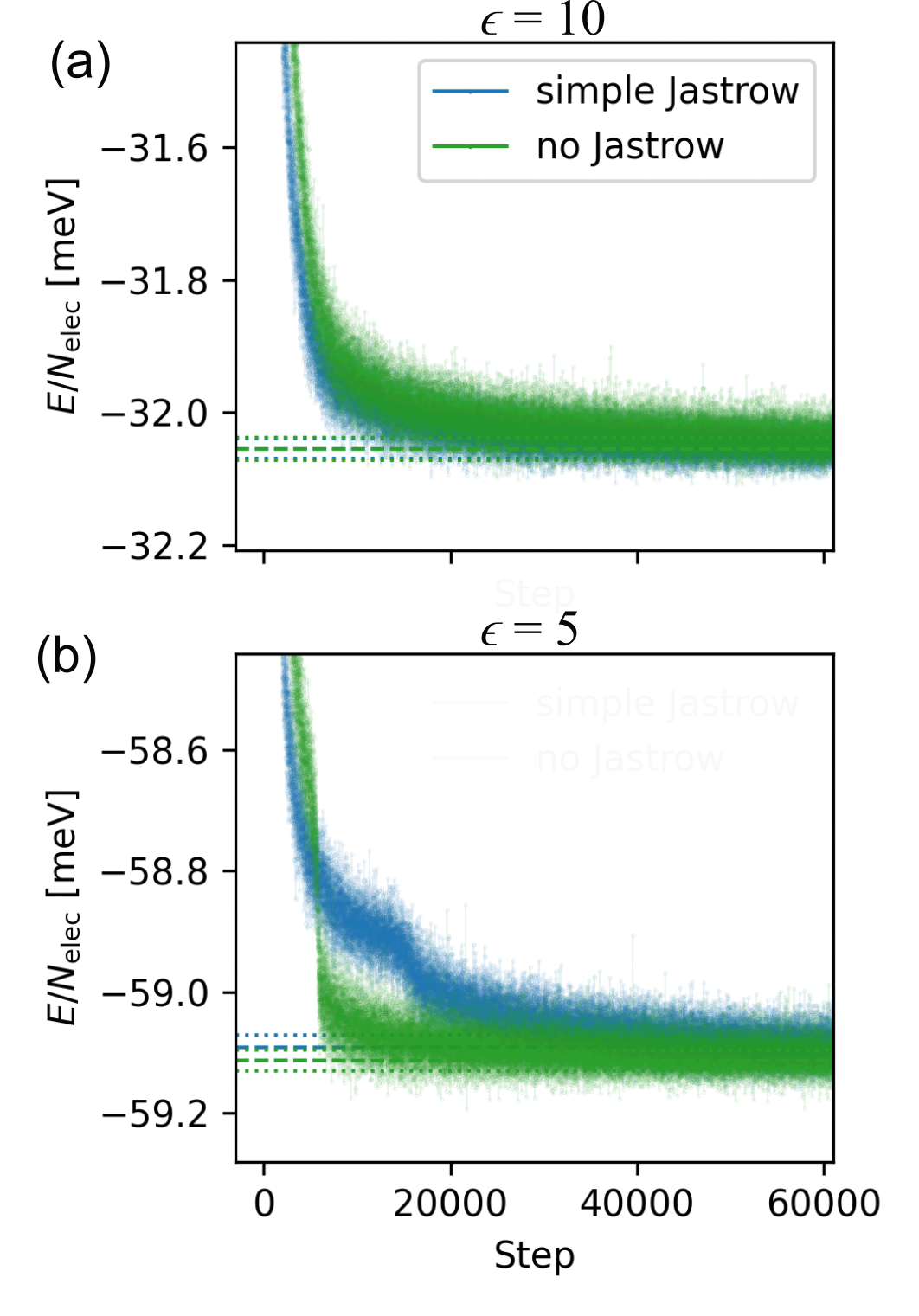}
    \caption{Comparison of training curves for 27 sites with 2/3 filling and (a) $\epsilon = 10$ and (b) $\epsilon = 5$ with simple Jastrow factor $J(\bm R)$ and $u(r)$ defined by Eq.~\eqref{eq:NN-cusp-u(r)} to fix cusp conditions (blue), and no Jastrow factor (green). Dashed and dotted lines indicate mean and standard deviation of the local energy averaged over batch size. These calculations were performed with batch size 1024. Each data point represents the local energy averaged over batch size at each step.}
    \label{fig:SM-jastrow-comparison}
\end{figure}

Concretely, adapting the derivation of Ref.~\cite{Foulkes2001Jan} to two dimensions, requiring that the local energy $E_{\rm loc}$ is finite for configurations with two particles approaching each other for the Coulomb gas Hamiltonian in dimensionless units [Eq.~\eqref{eq:system-hamiltonian}] leads to the conditions \cite{Teng2024Nov}
\begin{equation}
    \beta := \partial_r u(r) = \begin{cases}
        -1/3 & {\rm for\ parallel\  spins} \\
        -1 & {\rm for\ opposite\  spins} 
    \end{cases}
    \label{eq:NN-cusp-conditions}
\end{equation}
where $\partial_r$ is the partial derivative in the radial direction. The conditions for parallel and opposite spins are different because for the former, the Slater determinant part of the wavefunction vanishes while for the latter it remains finite as two particles approach each other. 

In Fig.~\ref{fig:SM-jastrow-comparison} we present numerical calculations including a simple Jastrow factor
\begin{equation}
    u(r) = - \beta \frac{\alpha^2}{\alpha + r}
    \label{eq:NN-cusp-u(r)}
\end{equation}
with a single learnable parameter $\alpha$ and $\beta$ determined by Eq.~\eqref{eq:NN-cusp-conditions} to enforce the analytically known behavior of $\Psi(\bm R)$ as two particles approach each other. 
In comparison to our calculations without Jastrow factor, the inclusion of the Jastrow factor did not significantly improve the results. Instead, we found faster training without the Jastrow factor.

In systems with periodic boundary condition, the electron-electron distance $r$ that enters the Jastrow factor is calculated with respect to a periodic and smooth norm that reproduces the Euclidean norm in the limit $r \to 0$. The smooth and periodic norm $||\bm r||_{M}$ with respect to the supercell vectors $M = (\bm L_1, \bm L_2)$ is defined as \cite{Cassella2023Jan}
\begin{align}
    ||\bm r||_{M} & = \frac{1}{2 \pi} \sqrt{\bm a^T M^T M\bm a + \bm b^T M^T M\bm b} \label{eq:NN-attn-jastrow-periodicnorm} \\
    \bm a & = 1 - \cos (G \bm r), \ \ \bm b  = \sin (G \bm r) \nonumber
\end{align}
where $G = 2\pi M^{-1}$ are the reciprocal supercell vectors.

\section{Hyperparameters}
\label{app:Hyperparameters} 

\begin{table}
    \centering
    \renewcommand{\arraystretch}{1.3}
    \begin{tabular}{ll l}
        \toprule
        & \textbf{Parameter} & \textbf{Value} \\
        \midrule
        \textbf{Architecture} & Network layers & $3$ \\
        & Attention heads per layer & $6$ \\
        & Attention dimension & $16$ \\
        & Perceptron dimension & $64$ \\
        & $\#$ perceptrons per layer & $1$ \\
        & Determinants & $4$\\
        \midrule
        \textbf{Training} & Training iterations & $15e4$ \\
        & Learning rate at time $t$ & $\eta_0(1 + \frac{t}{t_0})^{-1}$ \\
        & Initial learning rate $\eta_0$ & $10$ \\
        & Learning rate delay $t_0$ & $1e5$ \\
        & Local energy clipping $\rho$ & $5.0$ \\
        \midrule
        \textbf{MCMC} & Batch size & $4096$ \\
        \midrule
        \textbf{KFAC} & Norm constraint & $1e^{-3}$ \\
        & Damping & $1e^{-3}$ \\
        \bottomrule
    \end{tabular}
    \caption{Table of default hyperparameters used in our numerical calculations with the self-attention neural network.}
    \label{Tab:Hyperparams}
\end{table}

Table \ref{Tab:Hyperparams} presents the default hyperparameters used in our calculations, ensuring stable and monotonous convergence during training. As discussed in the main text, architectural parameters must be adjusted according to system size to achieve convergence. For the convergence and scaling law analysis, we explored various hyperparameter distributions. The table lists the set that successfully ensured convergence for the 18 electron system. These parameters are applied to generate the figures in the main text. 

Additionally, we note that batch size is a crucial factor in controlling the standard deviation of the energy curves, with error bars decreasing as $1/\sqrt{{\rm batch\  size}}$
 . In some cases, reducing the KFAC damping parameter resulted in slightly lower energies than those reported. However, we observed significant instabilities when training with reduced KFAC damping, often leading to corrupted training runs.

\begin{figure}
    \centering
    \includegraphics[width=\linewidth]{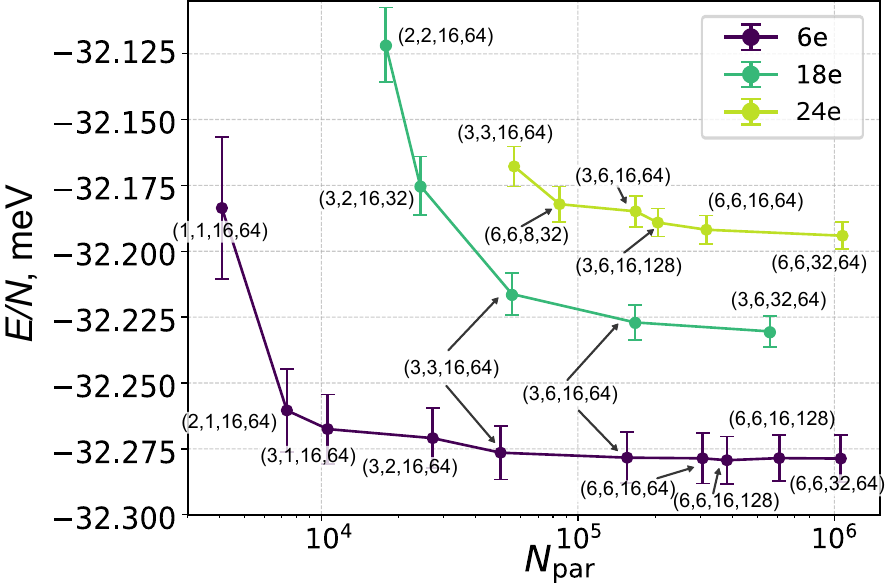}
    \caption{ Converged energy of the moir\'e system at $\nu=2/3$ filling and $\epsilon=10$ as a function of number of variational parameters. The labels indicate the parameters $(n_\text{layers}, n_\text{heads},  \text{dim}_\text{att}, \text{dim}_\text{perc})$ used for each calculation. The total number of parameters are counted explicitly from the checkpoints saved during the training. 
    }
    \label{fig:7_2}
\end{figure}

The Fig. \ref{fig:7_2} contains the precise network dimensions used to generate the parameter scaling in Fig.~\ref{fig:3} in the main text.

For the self-attention neural network as described in Sec.~\ref{sec:NN-attn}, the total number of variational parameters depends on the network dimensions as
\begin{align*}
    N_{\rm par} & = 2 d_{\rm dim} d_{\rm L} \\
    & + L \bigg(  2  N_{\rm heads} d_{\rm Attn} d_{\rm L}  \\ 
    & \qquad + 2  N_{\rm heads} d_{\rm AttnValues} d_{\rm L} \\ 
    & \qquad + d_{\rm L}( d_{\rm L} + 1 ) \bigg) \\
    & + 2 d_{\rm L} N_{\rm el} N_{\rm det} 
\end{align*}
 where the first line is the embedding $W_0$ of the periodic features in the layer dimension $d_L$, the second line are key and query matrices $W_{\rm k}^{lh},\ W_{\rm q}^{lh}$ for each layer $l = 1,...,L$ and head $h = 1, ..., N_{\rm heads}$, the third line are the value matrices $W_{\rm k}^{lh}$ and projection matrices $W_{\rm o}^{l}$ onto the layer dimension, the fourth line are the perceptron weights and biases $W^l, \bm b_l$, and the last line are the projection $\bm w_{2j}^m, \bm w_{2j+1}^m$ on $j = 1, ..., N_{\rm el}$ complex orbitals for $m = 1, ..., N_{\rm det}$ determinants.

\section{Correlation Energy}
\label{app:Correlation}

\begin{figure}
    \centering
    \includegraphics[width=0.8\linewidth]{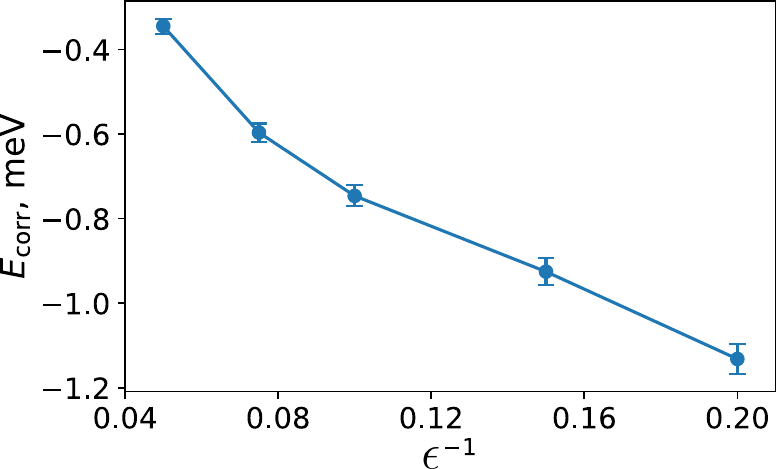}
    \caption{Correlation energy per electron, $E_\text{corr} = E_\text{self-attn NN}-E_\text{SlaterNet}$, at $\nu=2/3$ filling with $27$ sites as a function of inverse dielectric constant $\epsilon^{-1}$.  
    }
    \label{fig:8}
\end{figure}

Here, we investigate the correlation energy per electron, defined as the ground-state energy difference between the Hartree-Fock (SlaterNet) and \FNN approaches. We analyze its dependence on the interaction strength, controlled by the dielectric constant $\epsilon$, see Fig.~\ref{fig:8} for the result. As $\epsilon$ decreases (corresponding to stronger interactions), the correlation energy gradually increases, reflecting the growing importance of many-body effects beyond mean-field theory. This trend highlights the limitations of Hartree-Fock in capturing strong correlations and underscores the role of \FNN in accurately describing the correlated electronic states.

\section{BP-ED Benchmark}
\label{app:ED-benchmark}

\begin{figure}
    \centering
    \includegraphics[width=.99\linewidth]{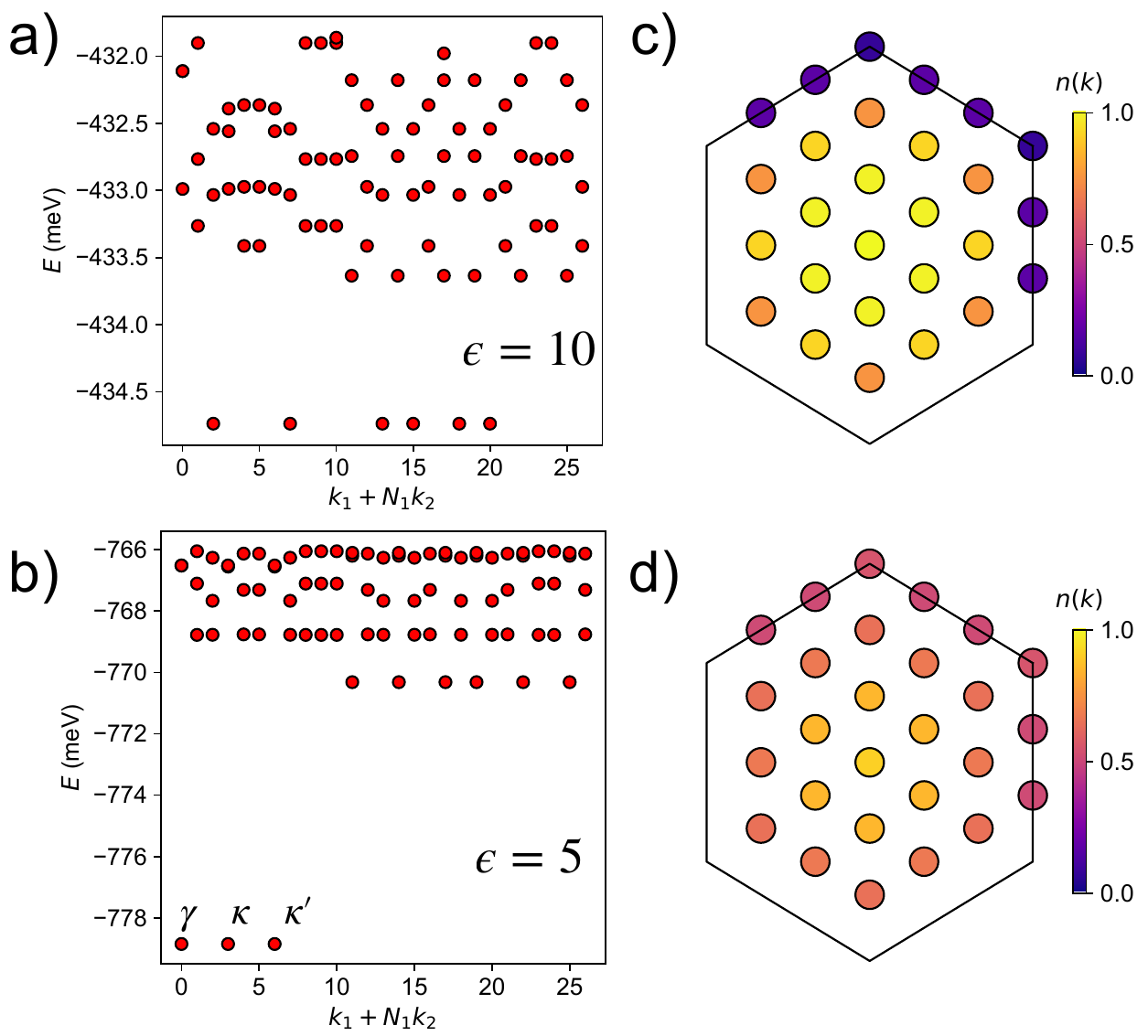}
    \caption{Many Body Spectrum (MBS) [(a),(b)] and ground state occupation [(c),(d)] for 27 sites at 2/3 filling for (above) $\epsilon=10$ and (below) $\epsilon=5$. The MBS of the left figure demonstrates the 6-fold GS degeneracy, while the MBS on the right demonstrates the 3-fold GS degeneracy at the high symmetry points, indicating a Generalized Wigner Crystal state. A clear Fermi Surface is also visible in (c) as the system is in a Fermi Liquid state.
    }
    \label{fig:9}
\end{figure}

In addition to benchmarking our \FNN energies with BP-ED, we leverage BP-ED to reveal deeper details on our system, namely identifying the Fermi-Liquid (FL) and Generalized Wigner Crystal (GWC) phases, along with the order and location of the metal-insulator (MIT) transition. Fig. \ref{fig:9} shows two distinct phase signatures at different interaction strengths $\epsilon=10$ [(a),(c)], and $\epsilon=5$ [(b),(d)]. By assigning each crystal momenta an integer index, the Many Body Spectra (MBS) in Fig. \ref{fig:9} ([a),(b)] shows the lowest energies per momentum sector, with the corresponding momentum space positions of each sector graphically displayed in the moiré Brillouin zone (MBZ) as a function of ground state (GS) occupation in Fig. \ref{fig:9} [(c),(d)]. 

The 6-fold rotational symmetry of the 27 site cluster allows the GS to accommodate a 6-fold degeneracy, emblematic of a high mobility, Fermi-Liquid-like state. On the other hand, the 27 site cluster also accommodates a tripled unit cell where the corners of the MBZ $\kappa, \kappa'$ fold back to $\gamma$ in the BZ, indicating an expected 3-fold GS degeneracy at $\gamma, \kappa, \kappa'$ for a GWC signature. Not only does the MBS in Fig \ref{fig:9} [(a),(b)] show clear signatures of a FL and GWC state at respective interactions, but also Fig \ref{fig:9} (c) shows the presence of a Fermi Surface while its $\epsilon^{-1}=5$ counterpart does not. 

\begin{figure}
    \centering
    \includegraphics[width=0.8\linewidth]{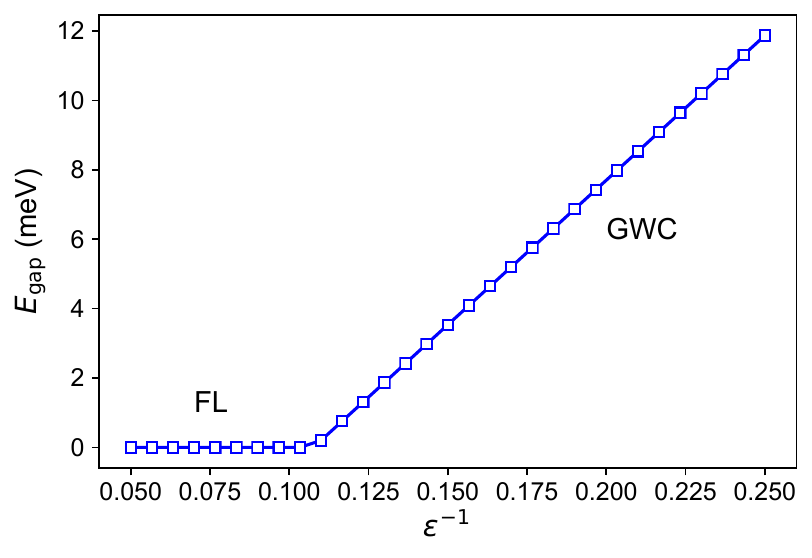}
    \caption{Many Body Energy Gap ($E_{gap}=E_4-E_3$, where $E_i$ labels the state with the $i^{th}$ lowest energy) as a function of interaction strength $\epsilon^{-1}$. The gap opens just after $\epsilon^{-1}=0.10$, signaling an MIT at about $\epsilon^{-1}=0.11$.
    }
    \label{fig:10}
\end{figure}

The exact MIT point predicted by BP-ED is shown in Fig. \ref{fig:10}, in which the many-body energy gap (difference between 4th and 3rd lowest energy values) are plotted versus interaction strength $\epsilon^{-1}$. Our expectation of an MIT as interaction strength increases is validated by the 6-fold GS degeneracy eventually giving way to a 3-fold GS degeneracy at about $\epsilon^{-1}=0.11$, signaling the crossover from FL to GWC. Our BP-ED results are well consistent with our \FNN results in Fig. \ref{fig:6}, as evidence for FL and GWC states are demonstrated at the appropriate interaction strengths, showcasing the ability of our \FNN to capture the physics of this moiré system.    

BP-ED techniques can also be used to gain insight on the nature of this MIT. We can characterize the transition as first or second order by probing the discontinuity of observables across the transition point. First order transitions are characterized by discontinuities across transition points while second order transitions are continuous across transition points for large enough system size. Since we are dealing with finite system sizes in BP-ED, there will always be a discontinuity; however if this discontinuity grows with system size, then we can deduce that correlation length will not tend to 0 with increasing system size, thereby showing evidence for a first order transition. Inversely, a decrease in the observable discontinuity across the transition point as system size grows is evidence for a second order phase transition. 

By probing the kinetic plus periodic potential one-body energy ($H_0$ in equation \ref{eq:system-hamiltonian}):
\begin{equation}
    \left\langle H_0 \right\rangle = \sum_{\mathbf{k}}\epsilon_{\mathbf{k}}\left\langle c_{\mathbf{k}}^\dagger c_{\mathbf{k}} \right\rangle_{\mathbf{GS}},
\end{equation}

and the band projected structure factor: 

\begin{equation}
    \Bar{S}(\mathbf{q})=\frac{1}{N_e}\left\langle \Bar{n}(\mathbf{q})\Bar{n}(\mathbf{-q})\right\rangle_{\mathbf{GS}},
\end{equation}

across the MIT point ($\epsilon^{-1}=0.11$) for 21 and 27 site clusters (21, 27-k-point mesh detailed in \cite{wilhelm2021interplay}), we can measure the discontinuity gap to shed light on the order of the phase transition. Since the projected structure factor has peaks along $\mathbf{q}=\kappa,\kappa'$, we plot $\Bar{S}(\mathbf{q}=\kappa)$ as a function of interaction strength to isolate the effect of the transition. 

\begin{figure}[b!]
    \centering
    \includegraphics[width=.99\linewidth]{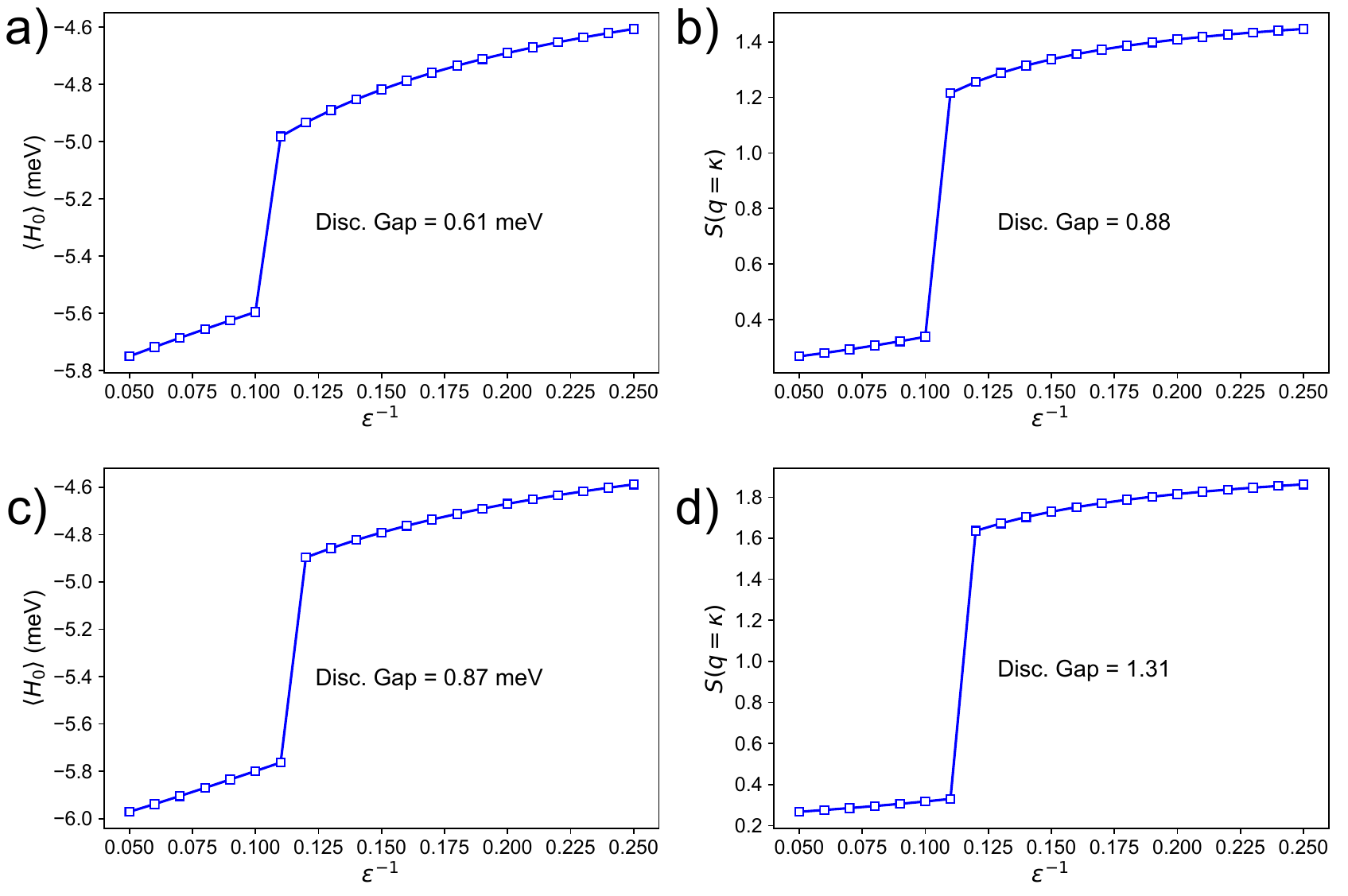}
    \caption{$\langle H_0 \rangle$ [(a),(c)] and $S(\mathbf{q}=\kappa)$ [(b),(d)] as a function of $\epsilon^{-1}$ for the 21 [(a-b)] and 27 site [(c-d)] systems. Here, $\langle ... \rangle_{GS}$ denotes the expectation value over a single ground state. The discontinuity gap increases with system size evincing a first order metal-insulator transition.
    }
    \label{fig:11}
\end{figure}

FIG. \ref{fig:11} not only demonstrates a discontinuity exactly at the critical point for each cluster, but also shows that the discontinuity gap grows as system size grows for both kinetic energy and the band projected structure factor (for $\mathbf{q}=\kappa$). The discontinuity gap for $\langle H_0 \rangle$ increases from 0.61 meV to 0.87 meV and from 0.88 to 1.35 for $\Bar{S}(\mathbf{q}=\kappa)$, demonstrating evidence for a first-order MIT. 

\end{appendix}

\bibliography{refs-MoireWC.bib}

\end{document}